\documentclass[a4paper,11pt]{article}
\pdfoutput=1 

\usepackage{jheppub} 

\usepackage[T1]{fontenc} 
\usepackage[multiple]{footmisc}
\usepackage{booktabs}
\usepackage{tabularx}
\usepackage{subfig}
\usepackage{snapshot}

\preprint{\parbox{3cm}{ZU-TH 18/14\\MCnet-14-09\\LPN14-068\\IPPP/14/34\\DCPT/14/68}}
\renewcommand{\arraystretch}{1.3}

\title{Standard Model Higgs boson pair production in the $(b\bar{b})(b\bar{b})$ final state}

\author[1,3]{Danilo Enoque Ferreira de Lima,}
\author[2]{Andreas Papaefstathiou,}
\author[3]{Michael Spannowsky}

\affiliation[1]{University of Glasgow, United Kingdom}
\affiliation[2]{Institut f\"ur Theoretische Physik, Universit\"at
  Z\"urich, Switzerland}
\affiliation[3]{Institute for Particle Physics Phenomenology, Durham
  University, United Kingdom}
\emailAdd{dferreir@mail.cern.ch}
\emailAdd{andreasp@physik.uzh.ch}
\emailAdd{michael.spannowsky@durham.ac.uk}

\abstract{Measuring the Higgs boson couplings as precisely as possible is one of the major goals of the High Luminosity LHC. We show
  that the $(b\bar{b})(b\bar{b})$ final state in Higgs boson
  pair production can be exploited in the boosted regime 
to give constraints on the trilinear Higgs
  boson self-coupling.
In these exclusive phase space regions, novel jet substructure techniques can be used to separate the signal from the large QCD and electroweak backgrounds. 
New developments on trigger and $b$-tagging strategies  for the upcoming LHC runs are necessary in order to reconstruct the Higgs bosons in boosted final states, where the trilinear self-coupling sensitivity is reduced.
We find that using our approach one can set a limit for $\lambda \leq 1.2$ at $95 \%$ CL after $3000~\mathrm{fb}^{-1}$. As the signal-to-background ratio is small, we propose a data-driven side-band analysis to improve on the coupling measurement.}

\begin{document} 
\maketitle
\flushbottom

\section{Introduction} 
The Standard  Model (SM) provides an exceptionally good description of the observed phenomena in the realm of elementary
particles, with very few exceptions, linked to much higher energy
scales that for the time being may lie beyond experimental
reach.\footnote{Examples of these may be the origin of neutrino masses
  or Grand Unification.} The epitomization of this success has been the discovery of
the final missing piece of the SM, the Higgs boson, by the CERN Large
Hadron Collider (LHC) experimental collaborations during the first run at
$pp$ centre-of-mass energies of 7
and 8~TeV~\cite{ATLAS_Higgs,
  CMS_Higgs, CMS-PAS-HIG-12-045, ATLAS:2012wma,
  ATLAS-CONF-2012-170}. The LHC's `Run II' is expected to
start at 13~TeV and potentially reach the nominal $pp$ energy of
14~TeV later on. Barring the exciting event of a phenomenal discovery
of new physics effects, the most important next step is to directly probe and
constrain the couplings of the Higgs boson to the content of the
SM. Of particular interest are the couplings of the Higgs boson to
itself, which will allow for understanding of the structure of the
symmetry breaking potential:
\begin{equation}
\mathcal{V} = \frac{1}{2}  M_h^2 h^2 + \lambda v  h^3 + \frac{\tilde{\lambda}}{4} h^4\;,
\end{equation}
where $h$ is the Higgs boson field, $M_h$ is the Higgs boson mass, $v$
is the vacuum expectation value, $\lambda$ and $\tilde{\lambda}$ are
the Higgs boson triple and quartic self-couplings respectively. 

The Higgs mass has already been determined to a good precision in the
first LHC run ($M_h \simeq 125$~GeV) and the vacuum expectation value,
$v \simeq 246$~GeV, has been obtained by measurements of four-fermion
interactions at low energies. These lead to the Standard Model
predictions: $\lambda = \tilde{\lambda} = M_h^2 / 2v^2 \simeq
0.13$.\footnote{Radiative corrections decrease these values by $\sim
  10\%$~\cite{Kanemura:2002vm,Grigo:2013rya}.} However, new physics can alter this direct correspondence and therefore, model-independently, the self-couplings $\lambda$ and $\tilde{\lambda}$ can be probed only by direct measurements of multiple Higgs boson final states. The quartic coupling,
$\tilde{\lambda}$ has been shown to be difficult, if not
impossible, to measure, even at future
colliders~\cite{Plehn:2005nk, Binoth:2006ym}. On the other hand, over the past 20
years, several studies have shown that the prospects for the Higgs
boson pair production, whose LHC cross section is a quadratic function
of the triple self-coupling $\lambda$, remain uncertain and
challenging both at future electron-positron collider
(e.g.~\cite{Miller:1999ct}) and at the LHC~\cite{Glover:1987nx, Dawson:1998py, Djouadi:1999rca, Plehn:1996wb, Baur:2002qd,
  Baur:2003gp, Dolan:2012rv, Baglio:2012np, Barr:2013tda, Dolan:2013rja,
  Papaefstathiou:2012qe, Goertz:2013kp, Goertz:2013eka,
  deFlorian:2013jea, deFlorian:2013uza, Grigo:2013rya, Cao:2013si,
  Gupta:2013zza, Nhung:2013lpa, Ellwanger:2013ova, No:2013wsa,
  McCullough:2013rea, Maierhofer:2013sha, Hollik:2001px,
  Dubinin:1998nt, Tian:2013yda, Dawson:2013bba, Lafaye:2000ec,
  Osland:1999ae,Osland:1999ad, Brucherseifer:2013qva, Yao:2013ika, Frederix:2014hta}.

Recent studies have demonstrated that using jet substructure
techniques~\cite{Seymour:1993mx} in the boosted regime can potentially provide a reasonable
constrain at the end of a high-luminosity run of the LHC~\cite{Dolan:2012rv,Baglio:2012np, Barr:2013tda, Dolan:2013rja,
  Papaefstathiou:2012qe, Goertz:2013kp,
  Goertz:2013eka, deFlorian:2013jea, deFlorian:2013uza, Grigo:2013rya,
  Cao:2013si, Gupta:2013zza, Nhung:2013lpa, Ellwanger:2013ova,
  No:2013wsa, McCullough:2013rea, Maierhofer:2013sha}, tackling the final states $hh \rightarrow (b\bar{b}) (\gamma
\gamma)$, $hh \rightarrow (b\bar{b}) (\tau^+
\tau^-)$ and $ hh \rightarrow (b\bar{b}) (W^+ W^-)$. The final state with the largest
branching ratio, $hh \rightarrow (b \bar{b})(b \bar{b})$ has also been 
considered in passing, and was deemed to be extremely
challenging \cite{Baur:2003gpa,Dolan:2012rv}, particularly if $\lambda \approx \lambda_{\mathrm{SM}}$. The difficulties can be
attributed to the enormous QCD background originating from multi-jet
production, of which the irreducible channel $b \bar{b}b \bar{b}$ is
also large, as well as the fact that the final state is fully hadronic
and thus challenging to trigger on.\footnote{Apart from measuring the
  Higgs boson self-coupling, in scenarios beyond the Standard Model, the di-Higgs final state can be significantly enhanced and the decay into $(b \bar{b})(b \bar{b})$ can become a promising discovery channel \cite{Dolan:2012ac, Gouzevitch:2013qca, No:2013wsa, Ellwanger:2013ova,Efrati:2014uta, ATLAS-CONF-2014-005, Cooper:2013kia, Han:2013sga}.}

As the search for the Higgs boson itself has shown, in the face of limited statistics
one can rely on the combination of multiple channels to obtain the
best constraints. This will always be the case in Higgs boson pair
production, as the total cross section for the SM value of the triple
coupling is estimated to be only $\sim
30-40~\mathrm{fb}$ when higher-order QCD corrections are included~\cite{Dawson:1998py, deFlorian:2013jea, deFlorian:2013uza, Shao:2013bz, Grigo:2013rya, Philippov:2006th, Dolgopolov:2003kv, Boudjema:2001ii}. Taking the results of Ref.~\cite{Goertz:2013kp} at
face value, one would get the constraints shown in Table~\ref{tb:constraints} from
each of the channels determined so far to be viable, given that the
self-coupling has the SM value, for an integrated LHC luminosity of
$3000$~fb$^{-1}$ at 14~TeV.\footnote{Note that due to the
increase in cross section as $\lambda$ decreases, the lower bound is
more stringent.} We emphasise the fact that constraints of this type are essentially
self-consistency tests of the SM: deviations from $\lambda =
\lambda_\mathrm{SM}$ would merely \textit{indicate} the presence of new effects.

On the other hand, in Ref.~\cite{Gupta:2013zza}, the magnitudes of deviations from the SM
value of the self-coupling in several explicit models of new physics were
estimated, given that no other dynamics associated with electroweak
symmetry breaking are seen. The conclusion was that, ideally, one would
like to detect deviations from $\lambda = \lambda_\mathrm{SM}$ that
are $\mathcal{O}(20\%)$ or less. To accomplish that level of accuracy,
one can for example supplement the current state of affairs with one or two channels which can constrain the
self-coupling at least equally as well as the $hh \rightarrow (b\bar{b}) (\tau^+
\tau^-)$ or $ hh \rightarrow (b\bar{b}) (W^+ W^-)$ channels at
3000~fb$^{-1}$. This argumentation illuminates the importance of opening up new search
channels to the determination of the Higgs boson self-coupling.
\begin{table}[t!]
  \begin{center}
    \begin{tabular}{@{}ll@{}}
      \toprule
      process & constraint ($\times \lambda_\mathrm{SM}$) \\ \midrule
       $hh \rightarrow (b\bar{b}) (\tau^+
\tau^-)$ & $\lambda = 1.00^{+0.40}_{-0.31}$ \\
       $hh \rightarrow (b\bar{b}) (\gamma \gamma)$ & $\lambda = 1.00^{+0.87}_{-0.52}$ \\ 
       $hh \rightarrow (b\bar{b}) (W^+ W^-)$ & $\lambda =
       1.00^{+0.46}_{-0.35}$ \\ \midrule
       combination & $\lambda = 1.00^{+0.35}_{-0.23}$ \\ \bottomrule
    \end{tabular}
  \end{center}
  \caption{The expected constraints for an integrated LHC luminosity of
    $3000$~fb$^{-1}$ (14~TeV), for each of the `viable' channels
    for Higgs boson pair production obtained by conservative estimates,
  according to Ref.~\cite{Goertz:2013kp}. The assumption used in obtaining
  these constraints is that the the self-coupling has the SM value. The final line provides the
  result originating from the
  naive combination in quadrature of these channels.}
\label{tb:constraints}
\end{table}

In the present article we demonstrate the possibility of using the $
pp \rightarrow hh \rightarrow (b\bar{b}) (b\bar{b})$ channel at the
LHC running at 14~TeV centre-of-mass energy, to constrain the
self-coupling by employing jet substructure techniques~\cite{Seymour:1993mx}, i.e. the so-called BDRS method~\cite{Butterworth:2008iy} and Shower Deconstruction \cite{Soper:2011cr, Soper:2012pb,Soper:2014rya}. While a variation of the former has already been used in this context in \cite{Dolan:2012rv}, here we perform a more detailed study complementing and combining the reconstruction using Shower Deconstruction.

The article is organised as follows: in Section~\ref{sec:pheno} we
describe some features of the kinematics of the Higgs boson pair
production process and provide more detail on the reconstruction methods used. In Section~\ref{sec:simanal} we provide details of the
Monte Carlo simulation for the signal and background and the analysis
strategy. In the same section we provide our results.
Concluding remarks are given in Section~\ref{sec:conclusions}. 

\section{Phenomenological considerations}\label{sec:pheno}
\subsection{Kinematics}
\begin{figure}[!htb]
  \begin{center}
    \vspace*{1ex}
    \includegraphics{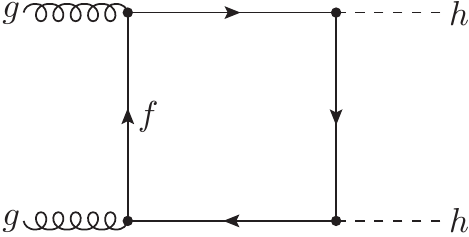}\qquad
    \includegraphics{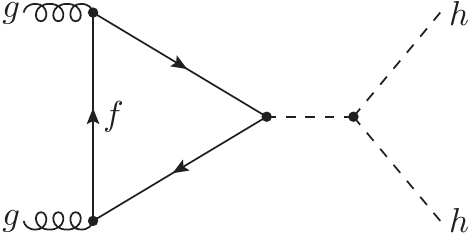}
    \vspace{-2ex}
  \end{center}
  \caption{Higgs boson pair production diagrams contributing to the
    gluon fusion process at LO are shown for a fermion $f$.
    These are generic diagrams and therefore, do not include all
    permutations.}
  \label{fig:HHdiagrams}
\end{figure} 
Higgs boson pair production at the LHC at leading order (LO) is
loop-initiated and dominated by gluon fusion
initial states. The contributing gluon fusion diagrams are shown in
Fig.~\ref{fig:HHdiagrams}. We call the diagram on the left the `box'
diagram and the diagram on the right the `triangle' diagram. The two
diagrams have spin-0 configurations of the initial state gluons
that interfere destructively. The box diagram also has a spin-2
configuration of the incoming gluons. The $hh$ cross section is a
quadratic function of $\lambda$ at LO, and hence possesses a minimum with
respect to it. This can be shown to lie around $\lambda \sim (2.4-2.5)\lambda_\mathrm{SM}$,
depending on the parton density functions (PDFs)
employed~\cite{Goertz:2013kp}. We will only examine values
on one side of the this minimum. It is natural to choose the lower
half, as it includes the SM value. For completeness, we also include
negative values of the self-coupling, and focus on the region $\lambda
\in \{-1.0, 2.4\}\times \lambda_\mathrm{SM}$. 

\begin{figure}[!htb]
  \begin{center}
    \vspace*{1ex}
    \includegraphics[scale=0.5]{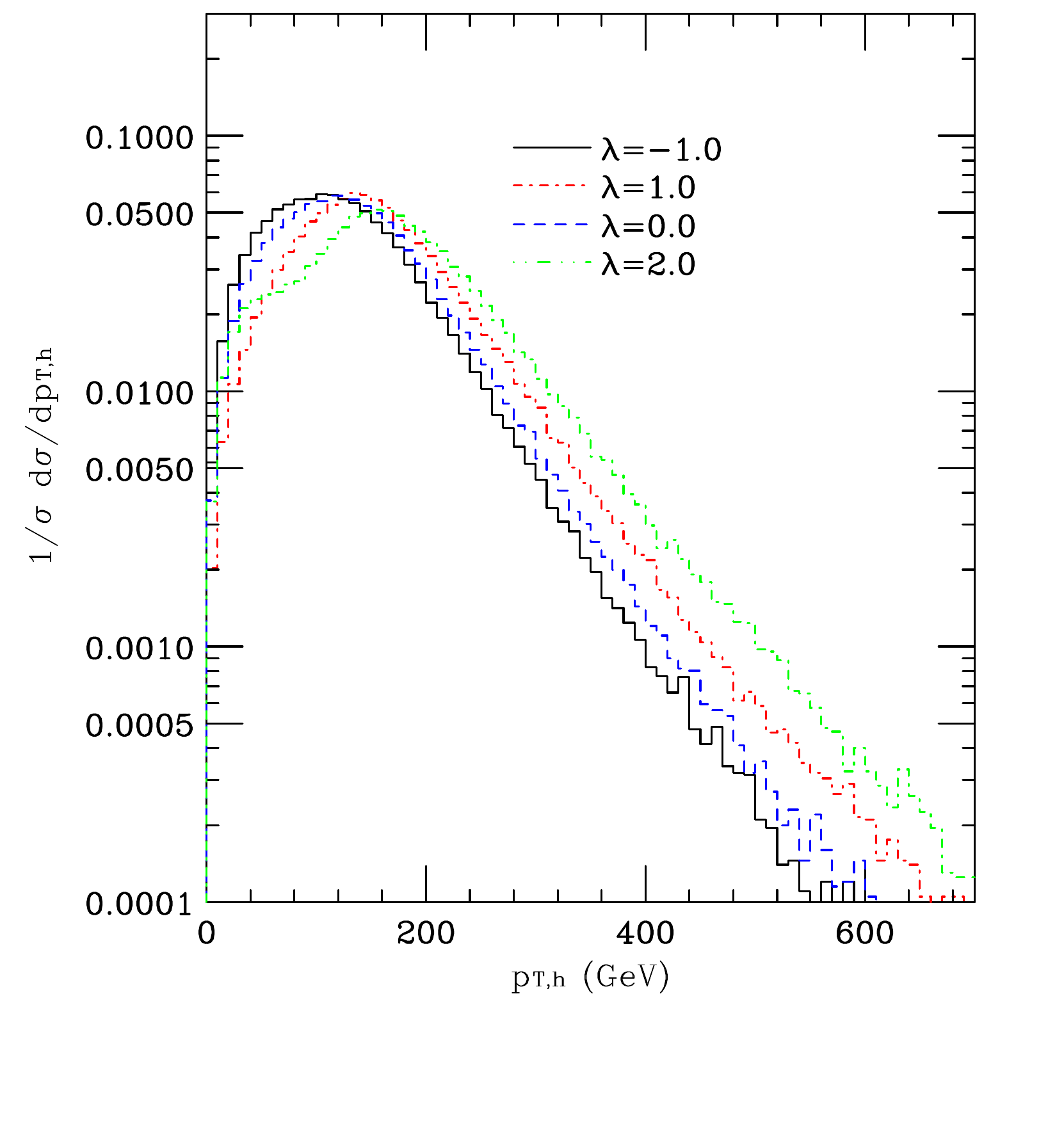}
    \vspace{-5ex}
  \end{center}
  \caption{The transverse momentum of a Higgs boson in the pair
    production process, including the box and triangle diagrams as well as their
interference, for several values of the self-coupling, given as
multiples of the SM value.}
  \label{fig:pth_lambdavar}
\end{figure}

It is interesting to examine the effect of varying the
self-coupling away from the SM value, on one of the characteristic distributions of the
process, namely, the Higgs boson transverse momentum. In
Fig.~\ref{fig:pth_lambdavar} we show the transverse momentum of the
full process, that is, including the box and triangle diagrams as well as their
interference, for several values of the self-coupling, given as
multiples of the SM value. Evidently, as $\lambda$ decreases the
distribution of the Higgs boson transverse momentum becomes
softer. This will result in a corresponding reduction of efficiency when a cut is applied on the transverse momentum of the reconstructed
Higgs boson. The dip structure observed in
Fig.~\ref{fig:pth_lambdavar}, prominent for $\lambda = 2\lambda_{SM}$, is a consequence of the destructive
interference between the box and triangle contributions.

\subsection{Event selection and triggering}
\label{subsec:evsel}

Triggering on events that contain purely hadronic final states is challenging.\footnote{We do not treat $b$-hadrons
  decaying into leptons in a separate way.} This is particularly so if the masses of the resonances involved are
at the electroweak scale and their decay products have transverse
momentum of only $\mathcal{O}(50)$ GeV. For the process $pp\to
hh \to  (b\bar{b})  (b\bar{b})$ that we are considering in the present
article, we have to rely on single-, di- or four-jet triggers.

\begin{figure}[!htb]
  \begin{center}
    \vspace*{1ex}
    \includegraphics[scale=0.45]{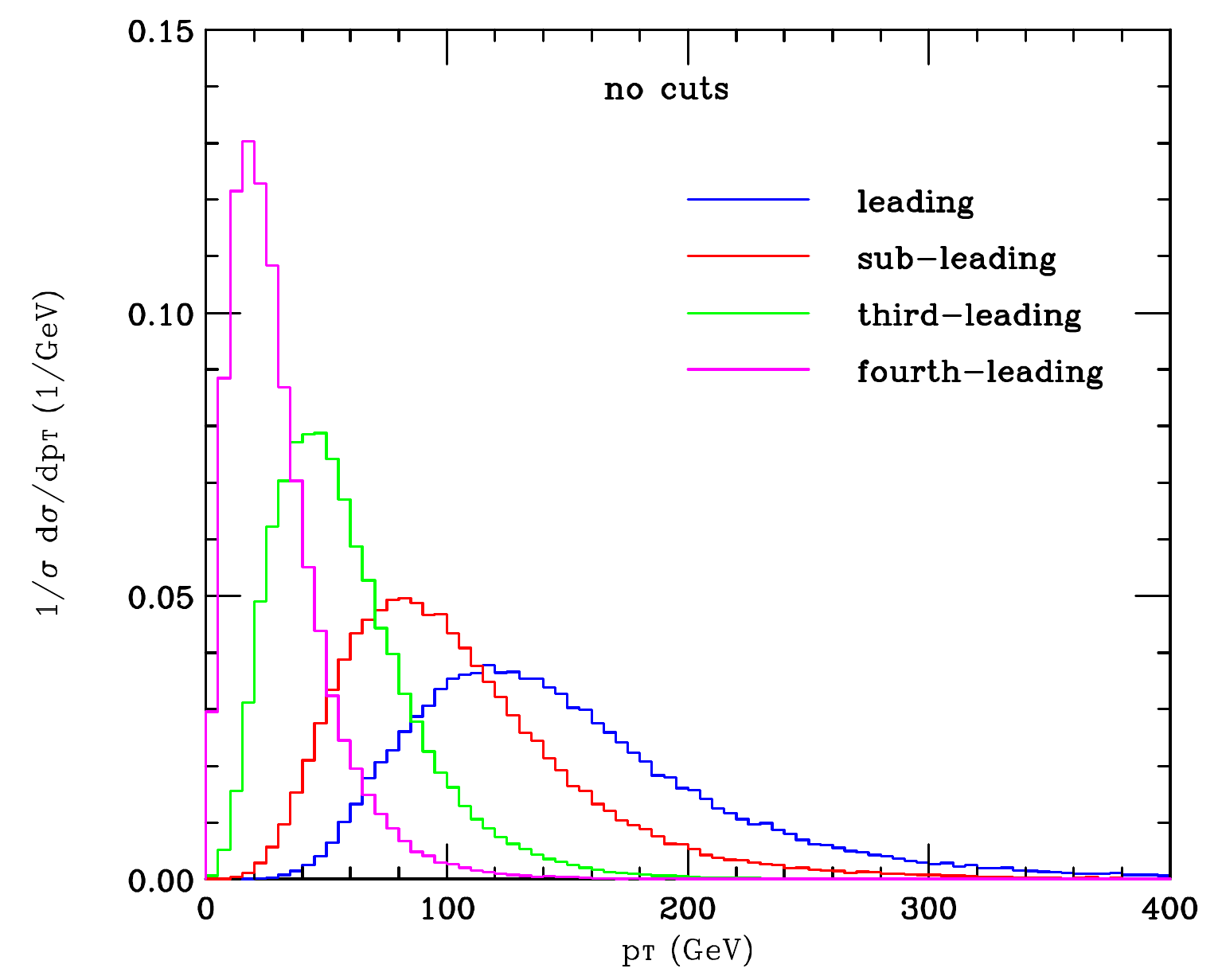}
    \includegraphics[scale=0.45]{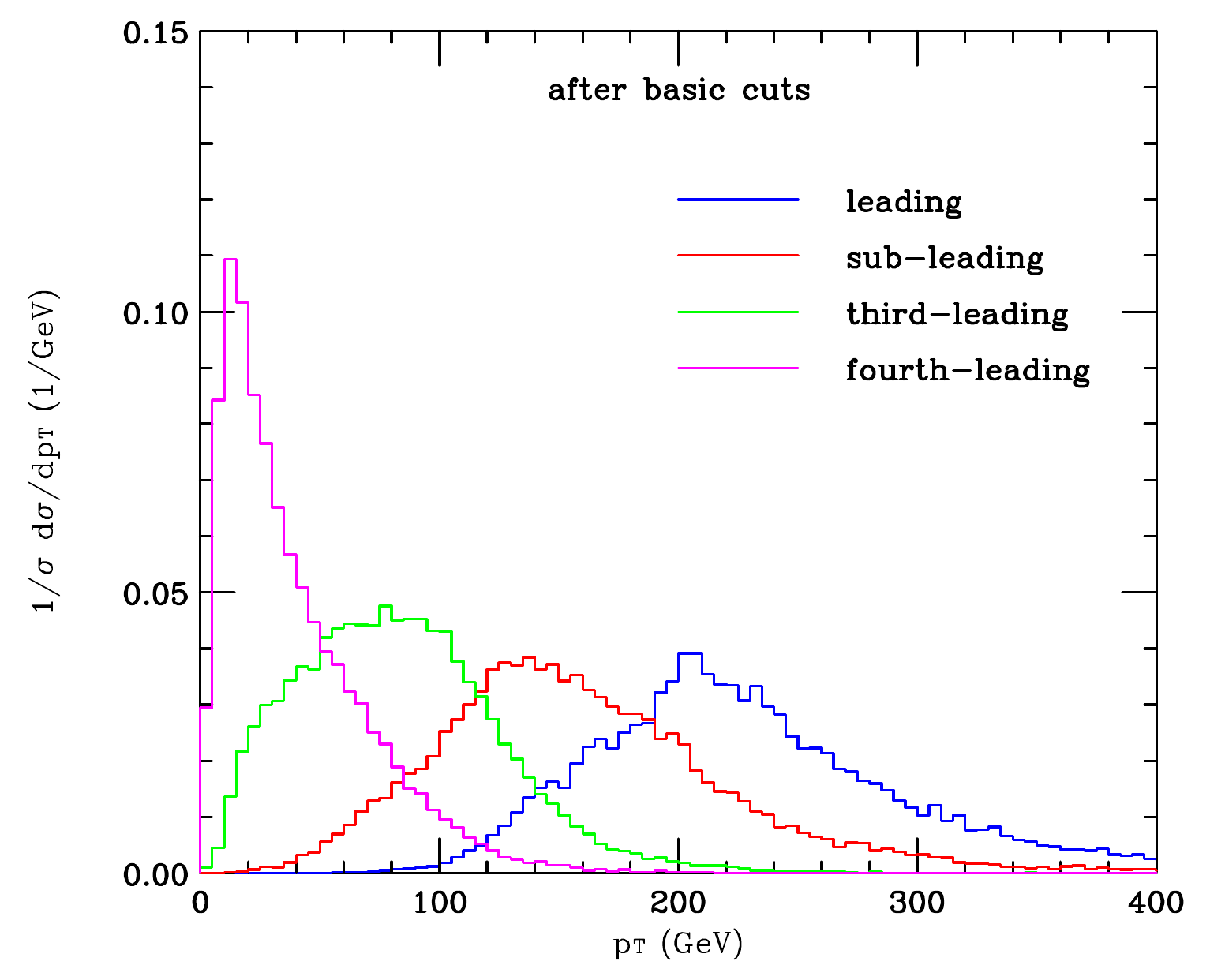}
    \vspace{-5ex}
  \end{center}
\caption{The transverse momentum of the four leading anti-$k_t$ $R=0.4$
b-jets in the SM $hh\rightarrow (b\bar{b}) (b\bar{b})$ signal, before any cuts
(left) and after the basic cuts as described in
Section~\ref{sec:basic} (right).}
  \label{fig:subjetpt}
\end{figure}

Four-jet triggers designed for the upcoming LHC runs are only fully efficient for anti-$k_T$ $R=0.4$ jets with transverse momenta of at least $60$ GeV~\cite{Bartoldus:1602235}.
The left panel of Fig.~\ref{fig:subjetpt} shows the transverse momentum of the four
leading $b$-jets in $pp\to
hh \to  (b\bar{b})  (b\bar{b})$ at an LHC running at 14~TeV, constructed using the anti-$k_t$ algorithm with
radius parameter $R=0.4$, before any cuts are applied. Less than
$10\%$ of the events have a fourth leading b-jet with $p_T \geq 60$
GeV. In some of those events a non-b-jet, e.g. a light hard jet
produced by initial state radiation, can help satisfy the trigger
requirements. However, it is evident that a large fraction of the
signal is lost only due to multi-jet trigger requirements.
Furthermore, using a radius parameter of $R=0.4$ would not allow to use jet substructure techniques to reconstruct boosted Higgs bosons efficiently.

Given that the Higgs bosons in the Standard Model $pp \to hh$ process
are often boosted transverse to the beam axis (see Fig.~\ref{fig:pth_lambdavar}) it seems to be more promising to rely on
a single-jet trigger with high transverse momentum~\cite{Bartoldus:1602235}. Thus, our basic event selection requires to have two  jets with a large radius parameter $R=1.2$, constructed with the 
Cambridge/Aachen jet algorithm. Each of these jets is required to
have a transverse momentum of $p_{T,j} \geq 200$~GeV. In this regime, the main
backgrounds can be substantially suppressed using jet substructure
techniques, as discussed in the following section. The Higgs boson
decay products should result in two $b$-jets contained within the
large-$R$ jet. Unfortunately, despite the fact that the large-$R$ jet is
fairly hard, the $b$-jets themselves can still possess a very low
transverse momentum: by examining the right panel of
Fig.~\ref{fig:subjetpt}, one notices that there remains a significant
fraction of signal in which the fourth $b$-jet has a low $p_T$, with a
substantial amount of events in the $p_T \sim 20$~GeV region, even
after the $p_{T,j} \geq 200$~GeV jets are selected. 

In developing the analysis of the present article, it was found that
when reconstructing the Higgs bosons of the process, one must require four $b$-tagged
subjets to facilitate background rejection. At present, good $b$-tagging performance with small systematic
uncertainties requires a jet transverse momentum of $30$-$40$ GeV \cite{ATLAS_btagging}. Aiming to evaluate what the sensitivity in this channel is with existing techniques we conservatively require the $b$-tagged subjets inside the fat jet to have $p_{T,b} > 40$ GeV. This results in a loss of signal events and seriously limits the effectiveness of novel subjet reconstruction techniques. In this analysis, four $b$-tagged jets are required in each event, assuming a 70\% $b$-tagging efficiency and a 1\% light jets false identification probability.

We stress that the angular jet resolution and transverse momentum
requirements for the multi-jet triggers, as well as the high momentum
cut-off for $b$-tagging of jets and subjets severely reduce the
sensitivity in this channel. Thus, an improved triggering strategy
would allow for much lower $p_T$ thresholds for the 4-jet system and
a smaller $R$ value for the jets in a real detector environment. A low-$p_T$ jet trigger with two or three $b$-jets in opposite hemispheres could help to regain sensitivity in this channel for the upcoming LHC runs.

\subsection{Reconstruction techniques}
\label{subsec:reconstruction}
To reconstruct the Higgs bosons in the $(b\bar{b}) (b\bar{b})$ final
state we use the well established BDRS method
\cite{Butterworth:2008iy} and Shower
Deconstruction~\cite{Soper:2011cr, Soper:2012pb,
  Soper:2014rya}.\footnote{Other reconstruction techniques could perform similarly~\cite{Thaler:2010tr, Soper:2010xk, Backovic:2012jj, Almeida:2011aa, Ellis:2012sn}.} Both methods aim to distinguish a jet that contains the decay of
products of a hadronically-decaying resonance from a jet produced by
ordinary QCD processes. The jet in question is constructed with a
standard jet algorithm, such as Cambridge/Aachen, using a large
radius parameter $R=1.2$ so as to capture the decay products of the heavy
resonance. This is what is referred to as the `fat jet'. 

For the method Shower Deconstruction the contents
of the fat jet are then used to construct narrower jets, which are called
`microjets'. Since the computational time needed to analyse an event
increases fast with the number of microjets, the number is restricted
by keeping the $N_\mathrm{max}$ microjets that have the highest
transverse momenta, and rejecting microjets with $p_T^\mathrm{micro} <
p_{T,\mathrm{min}}^\mathrm{micro}$. In what follows, we will employ
$N_\mathrm{max}=9$ and $p_{T,\mathrm{min}}^\mathrm{micro} = 5$~GeV. The four-momenta of the microjets $\{p\}_N = \{ p_1, p_2,
..., p_N\}$ are used to construct a function $\chi (\{p\}_N)$, with the
property that large $\chi$ corresponds to a high likelihood that the
jet has originated from the hadronic decay of a heavy
resonance. Explicitly, $\chi$ is defined as
\begin{equation}
\chi (\{p\}_N) = \frac{ P(\{p\}_N|S) }  { P(\{p\}_N|B) } \;,
\end{equation}
where $P(\{p\}_N|X)$ is the probability that the configuration
$\{p\}_N$ is obtained, given that the event originated from sample $X$,
where $X=S$, the signal, or $X=B$, the background. 
The probabilities defined above are calculated by using a simplified
approximation to how a Monte Carlo event generator constructs the
parton shower, as well as the decay of resonances. There are many
possible histories that could lead to a given configuration, and thus
one sums over the corresponding probabilities over all of them. A full
description of the approximations employed and further details on the method can be found in
Refs.~\cite{Soper:2011cr, Soper:2012pb}. 
\section{Simulation and analysis}\label{sec:simanal}
\subsection{Signal and backgrounds}
The signal process $pp \rightarrow hh \rightarrow (b\bar{b})
(b\bar{b})$ was generated using the \texttt{Herwig++} implementation of
Higgs boson pair production at LO~\cite{Bahr:2008pv, Arnold:2012fq, Bellm:2013lba}, using the cteq6l
LO PDF set. We varied the self-coupling in multiples of the Standard Model value in the region
$\lambda \in \{-1.0, 2.4\}\times \lambda_\mathrm{SM}$ in steps of
$\Delta \lambda = 0.1 \times \lambda_{SM}$. The cross
section at each point was calculated at next-to-leading order (NLO) in QCD, using
the \texttt{HPAIR} code~\cite{hpair, Plehn:1996wb, Dawson:1998py} and
the CT10nlo PDF sets. The black curve in Fig.~\ref{fig:sigma} shows
the variation of the total cross section against the Higgs boson
self-coupling, $\lambda$.\footnote{Note that the branching ratio for $hh \rightarrow (b\bar{b}) (b\bar{b})$, $\simeq
0.333$, has \textit{not} been applied to this curve.}

The irreducible QCD $b\bar{b} b \bar{b}$ background was generated using
\texttt{AlpGen}~\cite{Mangano:2002ea}, and passed to the \texttt{Herwig++} parton shower. The LO pdf
set cteq6l was again used. The renormalization/factorization scale for the calculation was set to the sum of the squared transverse masses of the $b$-quarks, i.e. $\mu ^2 = 4m_b ^2 + \sum_i p_{Ti} ^2$, where $p_{Ti}$ are the $b$-quark momenta and $m_b = 4.7$~GeV. We applied the parton-level generation cuts: $p_{Tb,\mathrm{min}} = 35$~GeV, $\Delta R_{\mathrm{min}}=0.1$,
$\eta_{b,\mathrm{max}} = 2.6$, resulting in a total tree-level cross
section of $\sigma_\mathrm{tree}(b\bar{b} b \bar{b}) \simeq 100$~pb. Using \texttt{MadGraph/aMC@NLO}~\cite{Frixione:2010ra,
  Alwall:2011uj, Alwall:2014hca}, with equivalent cuts and renormalization/factorization scale and the cteq6m NLO pdf set, the NLO $K$-factor was estimated to be $\sim 1.5$. Therefore, we apply a $K$-factor of $1.5$, resulting in a cross section $\sigma_\mathrm{NLO}(b\bar{b} b\bar{b}) \simeq 150$~pb.\footnote{See also~\cite{Binoth:2009rv,Greiner:2011mp} for details on NLO $b\bar{b} b \bar{b}$ production.}

Further irreducible backgrounds arise from production of a $Z$ boson
in association with a $b$-quark pair, $Zb\bar{b}$, and from associated Higgs-$Z$
boson production, $hZ$, with both the $Z$ and $h$ decaying to
$b\bar{b}$. For completeness, we also consider the reducible background coming from associated production of a Higgs boson with a
$W$ boson which subsequently decays to a charm and bottom
quark.\footnote{Note however that we do not simulate charm-jet to
  bottom-jet mis-tagging in our analysis. Inclusion of this effect
  will not alter the conclusions of our analysis, since the $hW$
  background is negligible.} The $Zb\bar{b}$ background was generated using
\texttt{MadGraph/aMC@NLO} at NLO in QCD, with $p_{Tb,\mathrm{min}} =
30$~GeV on the associated $b$-quarks. The $hZ$ and $hW$ backgrounds were likewise generated at NLO using \texttt{MadGraph/aMC@NLO}. The decays
of $Z$ and $h$ to $b\bar{b}$ were generated using \texttt{Herwig++}, without any
restriction imposed on the momentum of the $b$-quarks. Similarly, the
decay of a $W$ to a charm and a bottom quark was performed using
\texttt{Herwig++}. Including the
branching ratios, the NLO cross sections for these processes were
found to be $\sigma_\mathrm{NLO}(Zb\bar{b}) \simeq 8.8$~pb,
$\sigma_\mathrm{NLO}(hZ) \simeq 70$~fb and $\sigma_\mathrm{NLO}(hW)
\simeq 96$~fb. 

There are several additional QCD processes that may contribute as
reducible backgrounds due to mis-identification of light jets or
$c$-quark-initiated jets as $b$-jets. The most significant of these
are QCD $b\bar{b}c\bar{c}$, $b\bar{b}jj$, $c\bar{c}c\bar{c}$,
$c\bar{c}jj$ and multi-light-jet production. If we assume light-jet-to-$b$ and $c$-to-$b$ mis-tagging
probabilities of 1\% and 10\% respectively, the total tree-level cross
section contribution for the sum of these processes was estimated
using \texttt{AlpGen} to be $\sim 10$~pb. In terms of the kinematics of the jets, these are
expected to behave similarly to the reducible QCD $b\bar{b} b
\bar{b}$ that we consider in the present article in detail. Hence, even considering a large $K$-factor of
$\mathcal{O}(2)$, these processes would contribute to increase the
total QCD background cross section by $\mathcal{O}(20\%)$. For the
purposes of the present phenomenological study, we do not add this
contribution to the QCD background.

The second column in Table~\ref{tb:xsecs} demonstrates the initial cross
sections, $\sigma_\mathrm{initial}$, for the backgrounds considered, taking into account, where
relevant, the branching ratios, the generation-level cuts and the
applied $K$-factors. 
\subsection{Basic analysis}\label{sec:basic}
\begin{figure}[!htb]
  \begin{center}
    \vspace*{3ex}
    \includegraphics[scale=0.70]{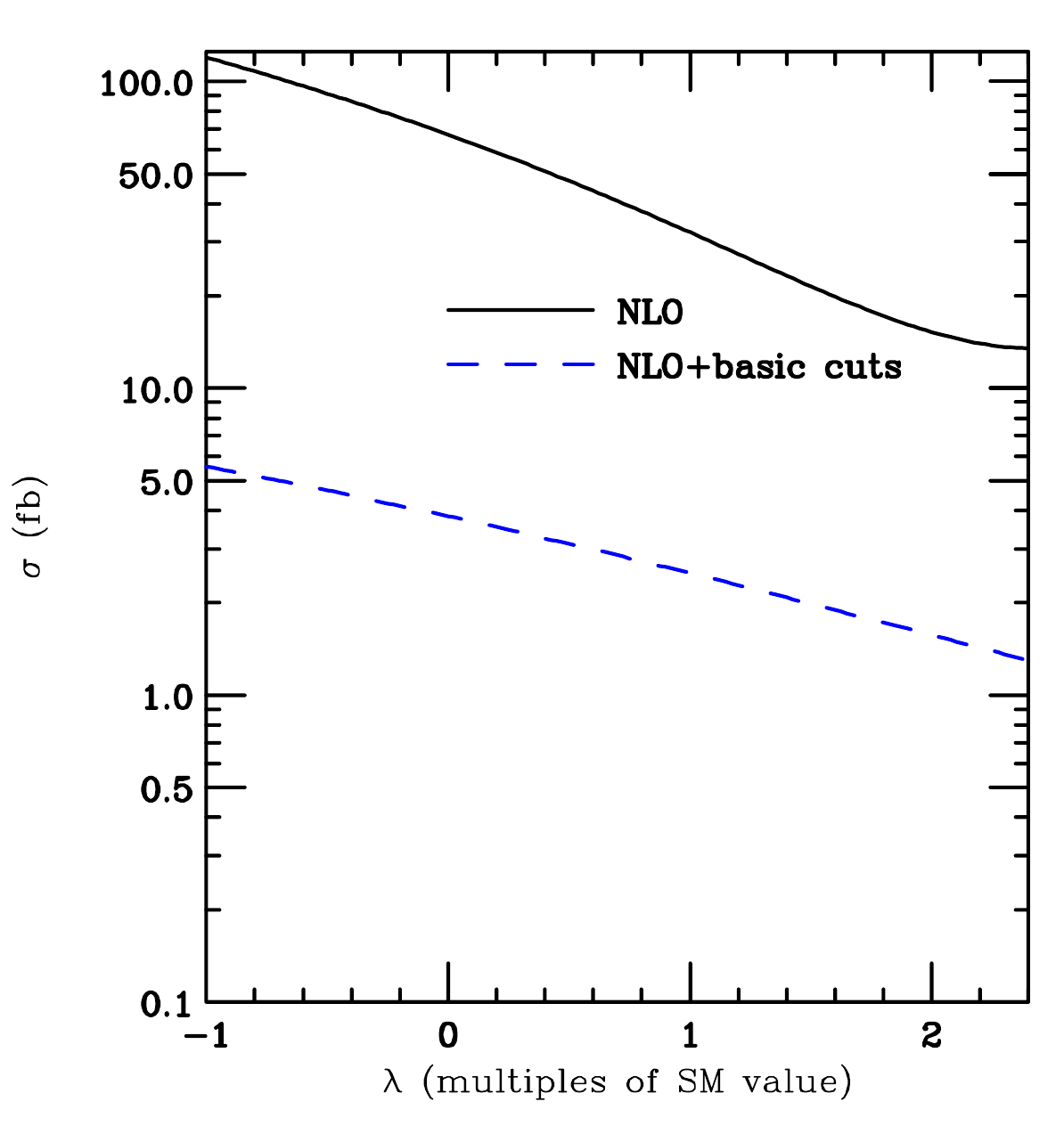}
    \vspace{-2ex}
  \end{center}
  \caption{The black curve shows the total NLO cross section at a
    14~TeV LHC calculated for
    each value of $\lambda$ using the \texttt{HPAIR} program, not
    including the branching ratio for $hh
  \rightarrow (b\bar{b})(b\bar{b})$. The blue dashed
  curve shows the resulting cross section after the `basic' analysis
  is applied to each signal sample, including the branching ratio for
  the $hh$ decays.}
  \label{fig:sigma}
\end{figure}

The basic analysis consists of the following simple cuts:
\begin{itemize}
\item{lepton isolation veto: Ask for \textit{no} isolated leptons with
    $p_T > 10$~GeV in the event. An isolated lepton is defined as having $\sum _i p_{T,i}$
  less than 10\% of its transverse momentum around a cone of $\Delta R
  = 0.3$ around it. }
\item{fat jets: ask for two fat jets built via the Cambridge/Aachen algorithm with parameter $R = 1.2$ and asking for $p_T > 200$~GeV.}
\end{itemize}
The resulting cross section as a function of $\lambda$, obtained after application of the basic analysis
cuts to the signal samples is shown in the blue dashed curve in
Fig.~\ref{fig:sigma}. The signal efficiency decreases with $\lambda$ and
hence the variation of the cross section in the region considered, after application of the
basic analysis cuts is somewhat milder than that of the total NLO cross
section: $\sigma^\mathrm{cuts}_\mathrm{NLO} \sim
1.3-5.6~\mathrm{fb}$ versus $\sigma_\mathrm{NLO} \sim
15-120~\mathrm{fb}$. The effect of the `basic analysis' on the cross
sections of all samples we have considered appear in the third column of
Table~\ref{tb:xsecs}. 
\begin{table}[t!]
  \begin{center}
    \begin{tabularx}{\linewidth}{XXX}
      \toprule
       sample & $\sigma_\mathrm{initial}$~(fb)
       &$\sigma_\mathrm{basic}$~(fb) \\ \midrule
       $hh$, $h\rightarrow b\bar{b}$ (SM) & 10.7 & 2.5 \\ 
       QCD $(b\bar{b})(b\bar{b})$ & 151.1$\times 10^3$ & 7.2$\times
       10^3$\\ 
       $Zb\bar{b}$, $Z\rightarrow b\bar{b}$ &8.8$\times 10^3$ & 284.2 \\
       $hZ$, $h\rightarrow b\bar{b}$,  $Z\rightarrow b\bar{b}$ & 70.0
       & 4.1 \\
       $hW$, $h\rightarrow b\bar{b}$, $W \rightarrow c\bar{b}
       (\bar{c}b)$ & 96.4 & 5.3 \\ \bottomrule 
    \end{tabularx}
  \end{center}
  \caption{The initial cross sections for the samples considered, as
    well as the resulting cross sections after cuts as described by
    the `basic' analysis. A $K$-factor of 1.5 was applied to the QCD
    $(b\bar{b})(b\bar{b})$ background. }
\label{tb:xsecs}
\end{table}

The basic cuts guarantee that the effect of the lepton veto is taken into
account in this analysis, which should minimise the effect of other
backgrounds with an isolated lepton in the final state. It also guarantees that the
$hh$ signal has a significantly high transverse momentum, which
already reduces a significant fraction of the QCD $b\bar{b}b\bar{b}$
background.
Nevertheless, the impact of the backgrounds is still very large,
as it can be seen in third column of Table~\ref{tb:xsecs}, and further signal-specific
constrains are necessary.

\begin{figure}[!htb]
  \begin{center}
    \vspace*{3ex}
    \includegraphics[width=0.7\linewidth]{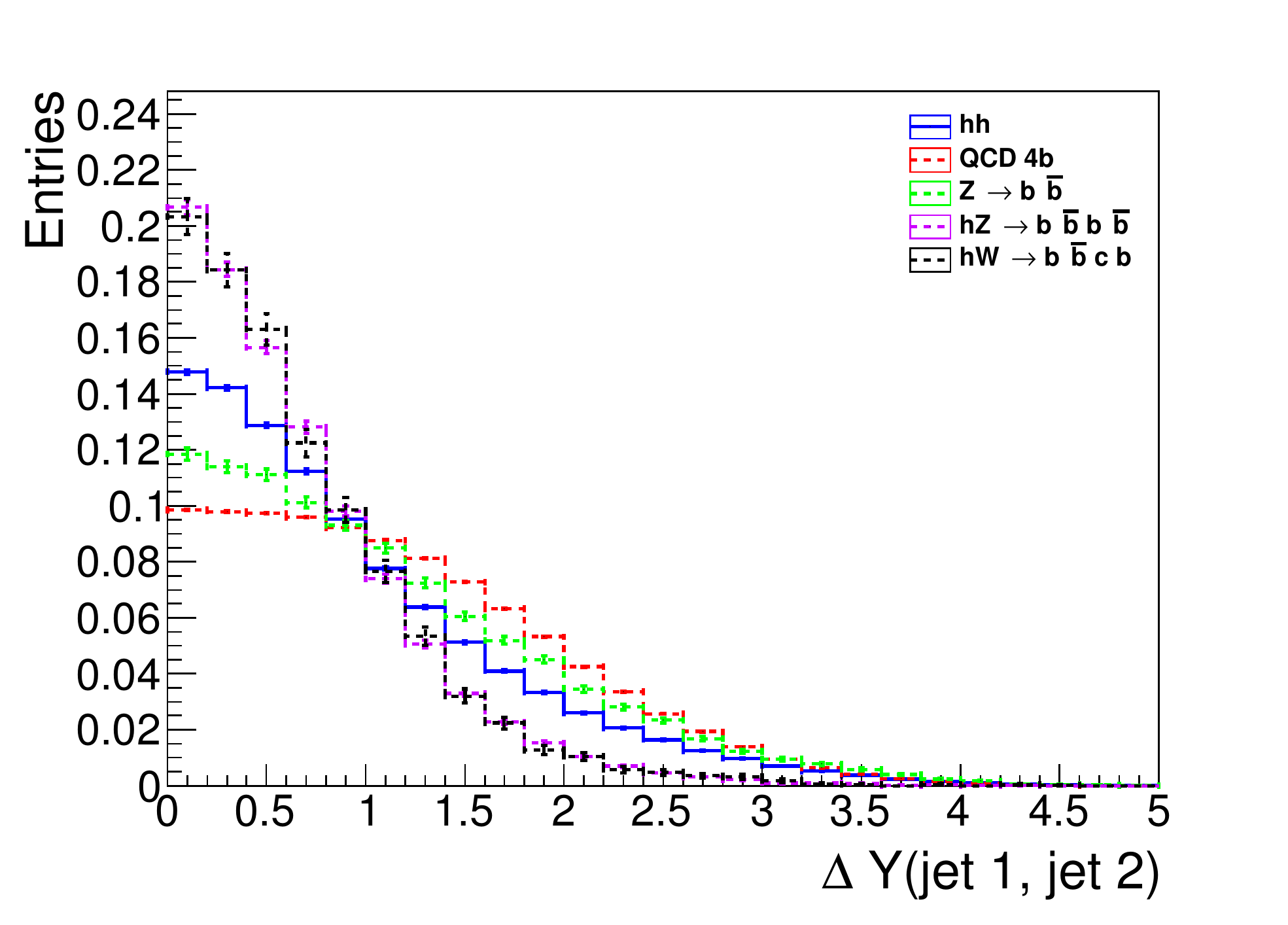}
    \vspace{-2ex}
  \end{center}
  \caption{The difference in rapidity between the two leading $R=1.2$ Cambridge-Aachen jets for the backgrounds and the $\lambda = \lambda_{SM}$ signal.}
  \label{fig:HHdeltaYlambda1}
\end{figure}

Figure~\ref{fig:HHdeltaYlambda1} shows the the distribution of the
difference between the rapidity values of the two large-$R$ jets,
$\Delta Y (\mathrm{jet~1},~\mathrm{jet~2})$, for various samples. All of the distributions
peak at $\Delta Y (\mathrm{jet~1},~\mathrm{jet~2}) = 0$, but evidently
the $hh$ signal has a narrower distribution compared to the
dominant $Zb\bar{b}$ and QCD $b\bar{b} b\bar{b}$ backgrounds.
A selection that takes only events with $\Delta Y
(\mathrm{jet~1},~\mathrm{jet~2}) < 2.0$ can therefore be beneficial to
increase background rejection.

Separating the $h\rightarrow b\bar{b}$ decay from the backgrounds can
be done using substructure techniques, either through Shower Deconstruction or
the BDRS reconstruction applied to the large-$R$ jet. The Shower Deconstruction~\cite{Soper:2011cr, Soper:2012pb}
Higgs boson tagger has been explored as a way of discriminating the signal
against the QCD multi-jet production, as described in
Section~\ref{subsec:reconstruction}.
The fat jet constituents
are used to calculate Cambridge/Aachen $R=0.2$ small-$R$ jets
(i.e. the microjets) which are
used as inputs for the algorithm. The leading three
microjets are examined and the ones with a transverse momentum of
at least $40$ GeV are required to pass $b$-tagging
criteria with a flat efficiency of 70\% and a false identification rate of
1\%. The Shower Deconstruction has been configured with a Higgs mass
window of $\pm 20$ GeV. The large $p_T$ cut for the $b$-tagged subjets
severely limits the performance of Shower Deconstruction as the
invariant mass of the two $b$-tagged subjets is often $\sim m_H$ leaving
not much phase space for wide-angle emissions off the bottom
quarks. Thus, the full flexibility of this algorithm is not exploited in this analysis.

Another approach uses the BDRS~\cite{Butterworth:2008iy}
method to reconstruct the Higgs boson
four-momentum. When using the BDRS,
a mass window can be applied to the reconstructed Higgs boson four-momentum
to minimize the background contamination.
A mass drop threshold of $\mu = 0.667$ and an
asymmetry requirement of $y_{\textrm{cut}} = 0.3$ are used.
As soon as a significant mass
drop is found in the fat jet, filtering is applied on the
jet's constituents, with a filtering radius value, $R_\mathrm{filt}$, of half the $k_t$ distance
between the mass drop elements. The filtering radius is limited to $R_\mathrm{filt} < 0.3$, following~\cite{Butterworth:2008iy}, but it is also
restricted to $R_\mathrm{filt} > 0.2$ so as to simulate of the impact
of the detector granularity limitation.
The three leading filtered jets
are taken as a result of this process to reconstruct the Higgs jet. The
jet is rejected however, if the two leading filtered
jets do not satisfy the
$b$-tagging criteria or have a transverse momentum below $40$ GeV.
The $b$-tagging criteria, as in the case of Shower Deconstruction, has 
70\% efficiency and a 1\% false identification rate.

\begin{figure}[!htb]
  \begin{center}
    \vspace*{3ex}
    \includegraphics[scale=0.70]{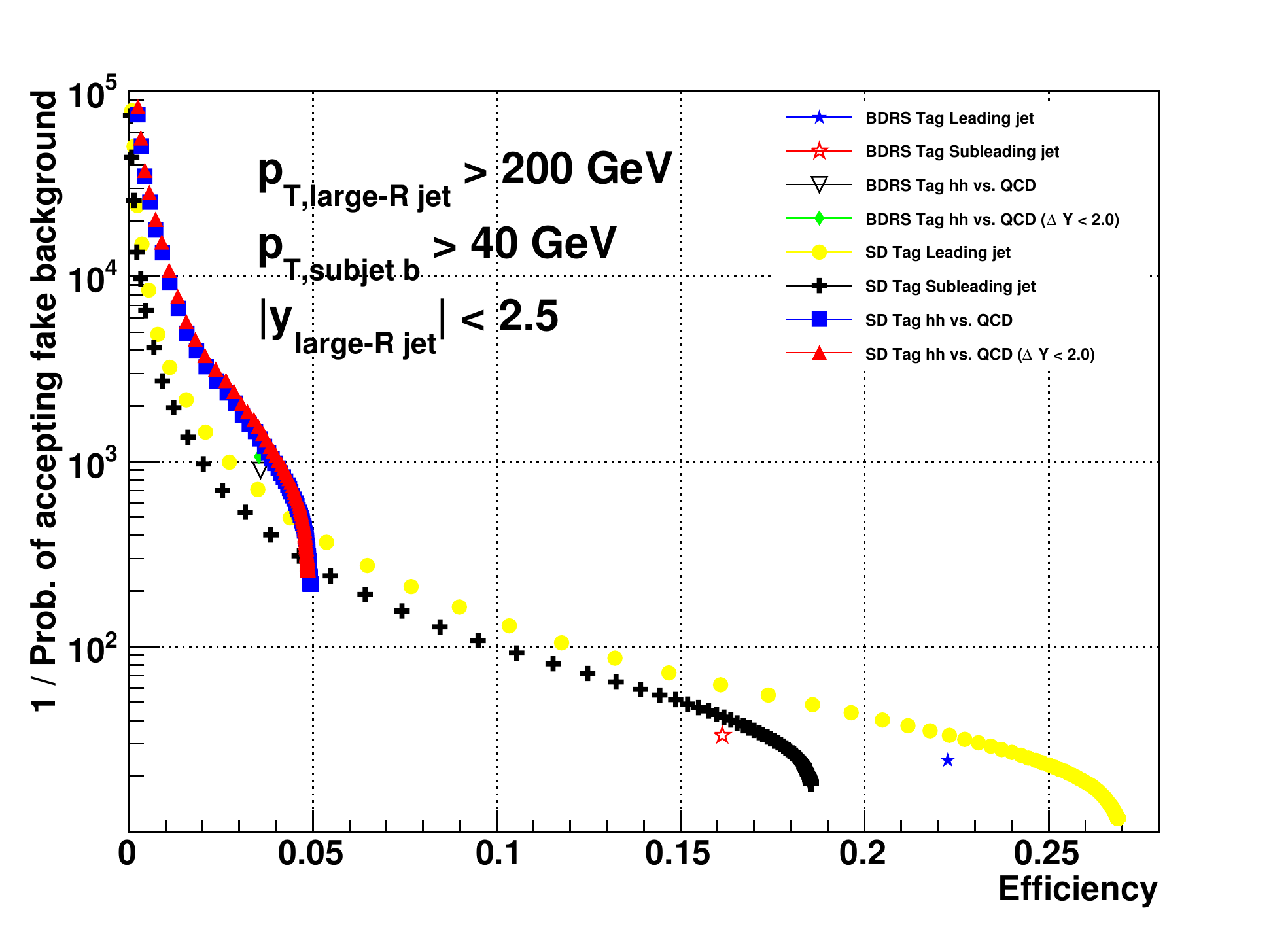}
    \vspace{-2ex}
  \end{center}
  \caption{The performance of the different Higgs boson-tagging
    methods for the process $pp \rightarrow hh \rightarrow (b\bar{b})(b\bar{b})$,
for $\lambda = \lambda_{SM}$. The efficiency includes two $b$-tags with $70\%$ tagging efficiency and $1\%$ fake rate and $p_{T,\mathrm{b}} \geq 40$ GeV for
the single jet tag lines, which is the cause of the limiting effect at $\sim 27\%$ efficiency. The lines showing the performance of tagging two Higgs bosons, include
the same effects for both jets.}
  \label{fig:roclambda1}
\end{figure}

Figure~\ref{fig:roclambda1} shows the performance of the different
`Higgs boson-tagging' methods we considered in the present article,
for the process $pp \rightarrow hh \rightarrow (b\bar{b})(b\bar{b})$,
for $\lambda = \lambda_{SM}$. The performance of the BDRS Higgs boson tagger with a mass cut of $\pm 20$ GeV
around the Higgs boson mass is shown, for which
the blue star marker (efficiency~$\sim$~0.22) shows the method applied to the leading jet only and
the red unfilled star marker (efficiency~$\sim$~0.16) shows the performance for the sub-leading jet.
The unfilled black downwards-pointing triangle (efficiency~$\sim$~0.04) shows the performance
of simultaneously applying the mass window to both leading and sub-leading jets,
while the green filled rhomboid (slightly above) has a further restriction on the
rapidity difference between the two jets at $\Delta Y (\mathrm{jet~1},~\mathrm{jet~2}) < 2$.
The equivalent performance for the Shower Deconstruction Higgs tagger
is shown for the same cases in the yellow circle, the black cross,
blue square and red up triangle, respectively.
Note that the efficiency axis is limited to $\sim 0.27$, due to the
subjets' $b$-tagging performance,
including the $40$ GeV threshold for $b$-tagged
jets.

\begin{figure}[!htb]
  \begin{center}
    \vspace*{3ex}
    \subfloat[Signal-to-background ratio.]{\includegraphics[width=0.48\linewidth]{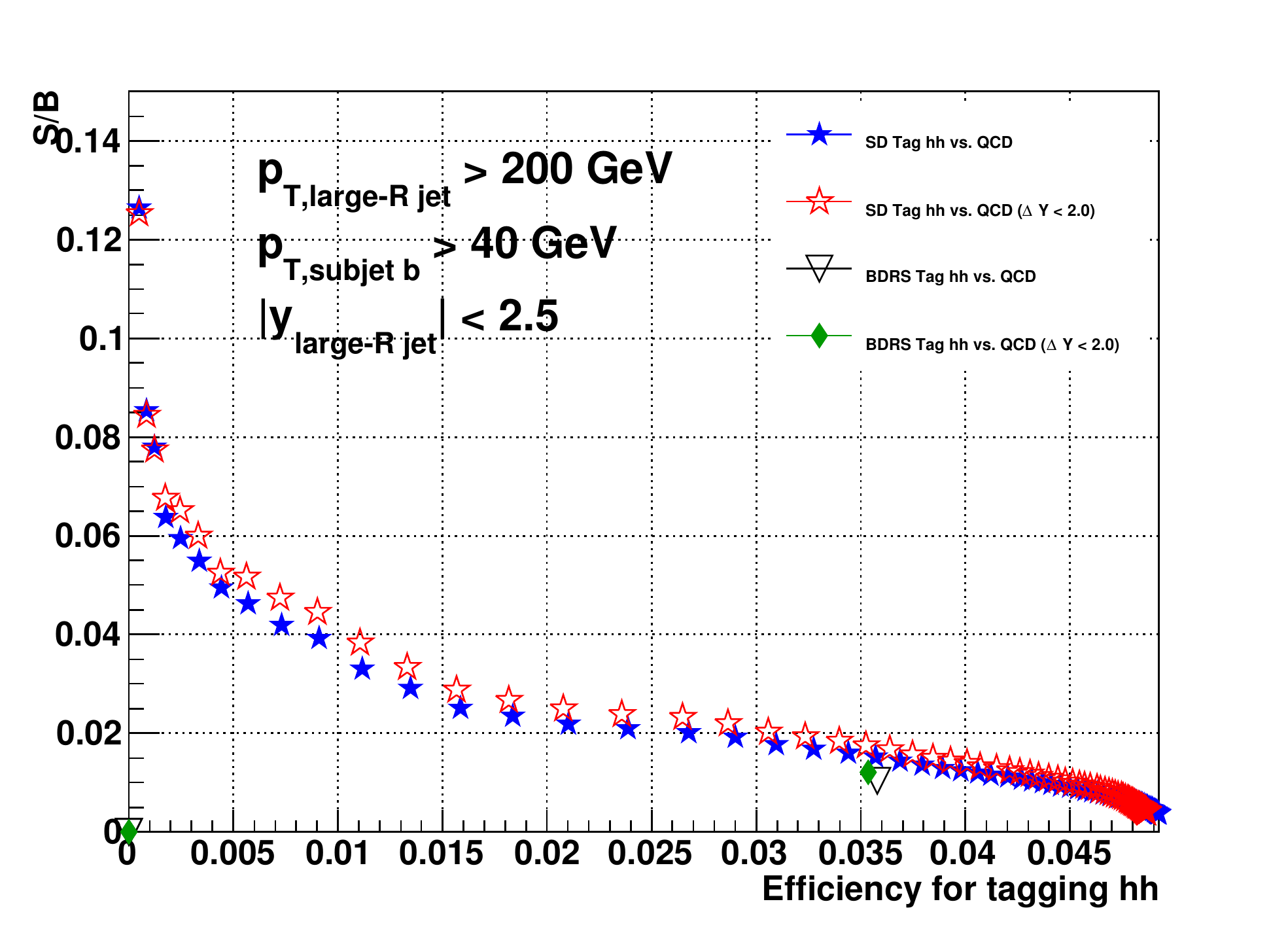}}
    \subfloat[Signal to square root of background ratio.]{\includegraphics[width=0.48\linewidth]{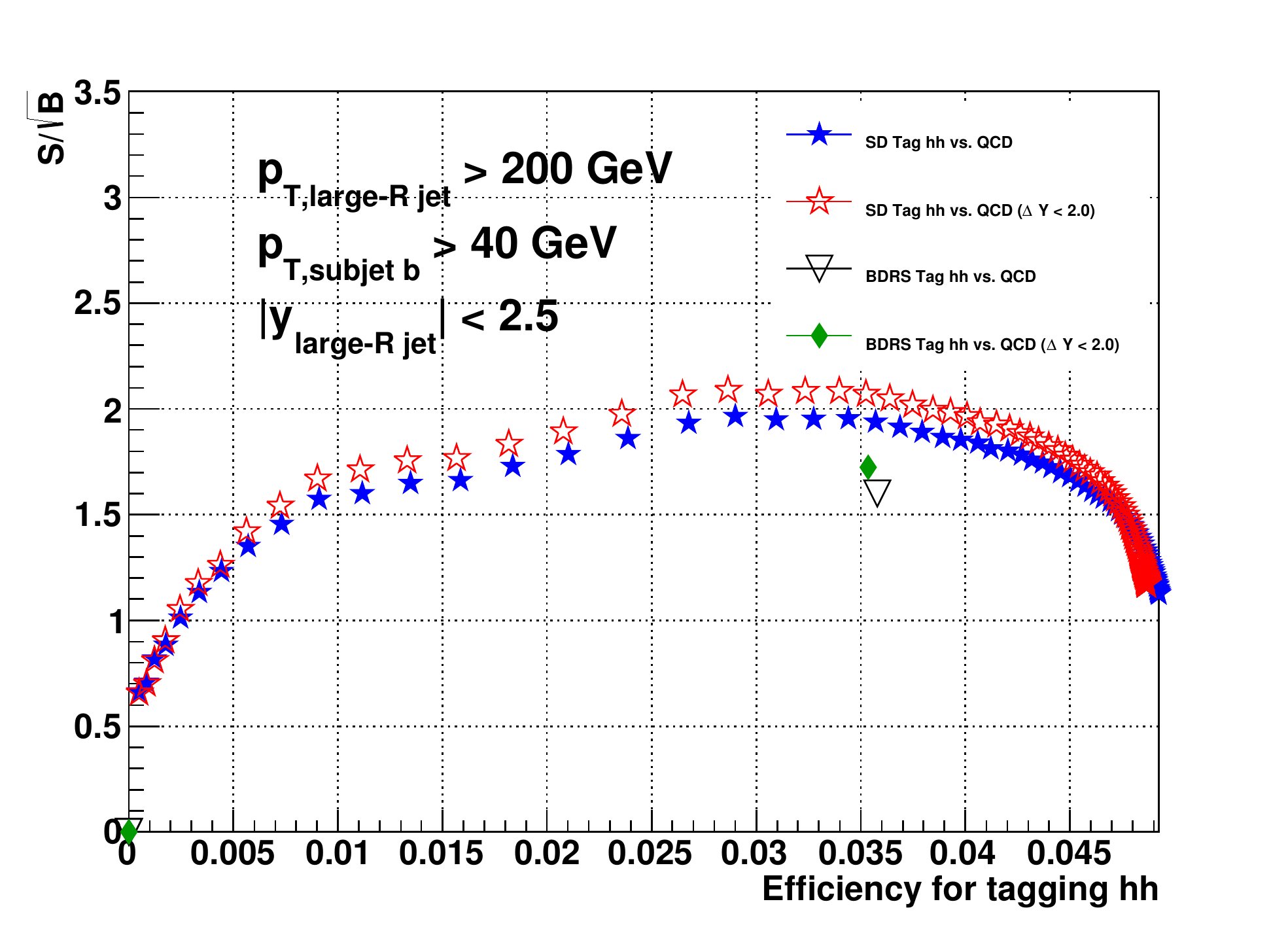}}
    \vspace{-2ex}
  \end{center}
  \caption{The signal-to-background ratio and expected significance of
    the different Higgs boson tagger methods for $\lambda = \lambda_{SM}$.}
  \label{fig:siglambda1}
\end{figure}

As we already hinted, the large cut on the jet transverse momentum reduces the
effectiveness of the Shower Deconstruction method, which ends up
providing an improved, but similar, background rejection as the BDRS
method at equal efficiency. Nevertheless,
the Shower Deconstruction technique allows one to achieve a high
rejection of the backgrounds, by varying the minimum weight requirement.
Figure~\ref{fig:siglambda1} shows the signal-to-background ratio and the significance estimator $s/\sqrt{b}$ for $\lambda = \lambda_{SM}$. The additional rejection given by
Shower Deconstruction allows one to have an increased signal-to-background
ratio, while maintaining a high value of the significance.

\subsection{Constraints}

Choosing the Shower Deconstruction setup that yields the maximum value of $s/\sqrt{b}$
from Fig.~\ref{fig:siglambda1}, one can estimate the maximum
significance an analysis would obtain with the described cuts.
Table~\ref{tab:yields} shows the cross sections obtained with this
selection, either using Shower Deconstruction only or BDRS only
for the Higgs jet tag.
The first row shows the cross sections after demanding two Cambridge/Aachen
jets with $R=1.2$ and transverse momentum grater than $200$ GeV, with
the additional constraint of rapidity $|y| < 2.5$, without the $b$-tagging
criteria.\footnote{Note that without the requirement of four $b$-tags, it cannot be
guaranteed that that the shown backgrounds are the dominant ones in
this case. Therefore the $s/b$ and $s/\sqrt{b}$ values shown are not
realistic and only shown for completeness.}

The following two rows show the effect of applying either Shower
Deconstruction or BDRS only to the leading fat jet, including
the effect of the two $b$-tags in its subjets. It also
includes the requirement
that the two large-$R$ jets satisfy $\Delta Y
(\mathrm{jet~1},~\mathrm{jet~2}) < 2.0$. The next two rows apply the BDRS or Shower Deconstruction requirements
to both jets, showing the final significance obtained of $\sim 2.10$ using
Shower Deconstruction, or $\sim 1.74$ using only the BDRS technique at
3000~fb$^{-1}$.

The last two rows in Table~\ref{tab:yields} show the significance achieved
by using a very loose Shower Deconstruction setting with the $26\%$
efficiency point of Fig.~\ref{fig:roclambda1} and using the BDRS
for the sub-leading jet reconstruction. The last row shows the effect of
applying a further mass window for the sub-leading jet. This selection
configuration
can be used to maximise the background in an attempt to implement
a data-driven background estimate through a side band analysis,
as it will be discussed in Section~\ref{subsec:sideband}.

\begin{table}[ht]
\footnotesize
\begin{tabular}{cccccccc}
\toprule
Selection  & hh & QCD 4b & hW & Z $\rightarrow$ bb & hZ & $s/b$ & $s/\sqrt{b}$\\
\toprule
Event selection  & 2.31 & 6941.860 & 4.854 & 266.472 & 3.787 & 0.000320 & 1.48\\
\hline
Leading jet SD  & 0.514 & 208.728 & 0.587 & 5.360 & 0.439912 & 0.00239 & 1.919\\
Leading jet BDRS  & 0.0982 & 54.223 & 0.0117 & 0.741 & 0.123 & 0.00178 & 0.724\\
\hline
Both SD-tags  & 0.0784 & 4.226 & $<$ 0.00096 & 0.0294 & 0.00605 & 0.0184 & 2.082\\
Both BDRS-tags  & 0.0817 & 6.671 & 0.000192 & 0.0593 & 0.00946 & 0.0121 & 1.723\\
\hline
Loose SD and BDRS rec. & 0.621 & 592.145 & 0.686 & 17.228 & 0.627 & 0.00101 & 1.376\\
Loose SD and BDRS & 0.0989 & 17.080 & 0.000612 & 0.129 & 0.0231 & 0.00574 & 1.305\\
\bottomrule 
\end{tabular}
\caption{Expected cross sections after selection for
  $\lambda/\lambda_{SM} = 1$. Note that the first row differs from
  $\sigma_\mathrm{basic}$ given in Table~\ref{tb:xsecs} due to the additional constraint on the
  rapidity of the fat jets, $|y| < 2.5$. The significance estimate, $s/\sqrt{b}$,
  given for an integrated luminosity of 3000 fb$^{-1}$. The two final rows show the
  results obtained using Shower Deconstruction on the leading jet and the
  BDRS for the Higgs reconstruction on the sub-leading one. In the
  last row a final mass cut on the sub-leading Higgs mass is applied.}
\label{tab:yields}
\end{table}

\begin{figure}[!htb]
  \begin{center}
    \vspace*{3ex}
    \includegraphics[width=0.7\linewidth]{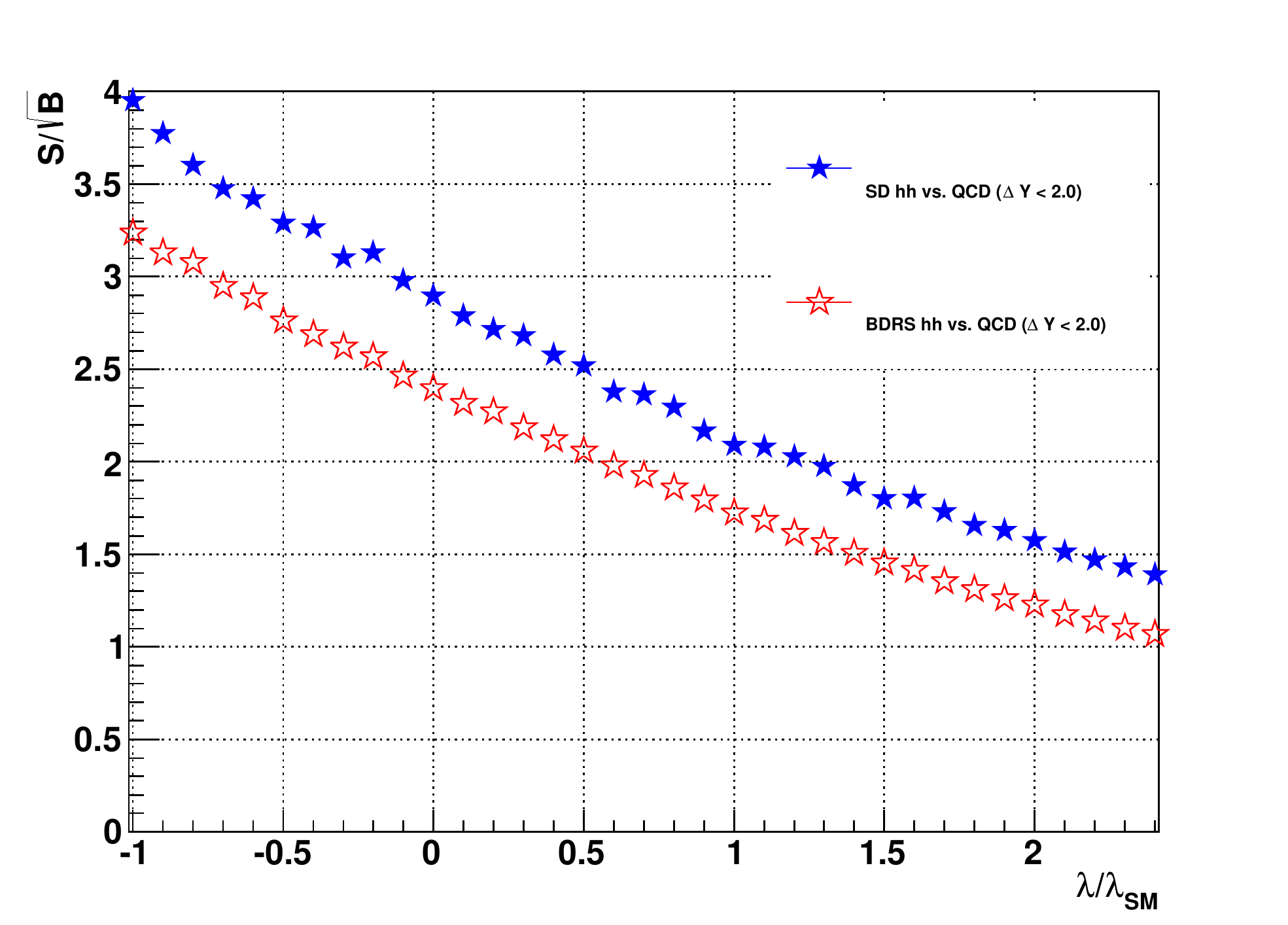}
    \vspace{-2ex}
  \end{center}
  \caption{The best expected significance of the different Higgs
    tagger methods for different values of $\lambda$ at 3000~fb$^{-1}$
    for a 14~TeV LHC.}
  \label{fig:sig_per_lambda}
\end{figure}

The analysis procedure can be applied at different values of the Higgs boson
self-coupling, and one can take the maximum achievable significance
obtained using Shower Deconstruction for each, at 3000~fb$^{-1}$. The results are shown in
Fig.~\ref{fig:sig_per_lambda}. Evidently, for the
Standard Model value of the self-coupling, $\lambda/\lambda_{SM} = 1$, it is
possible to set a 95\% limit using Shower Deconstruction. Due to the
reduction in cross section, the
significance drops as $\lambda$ increases. However, the hypothesis of
$\lambda = 0$, for example, can be excluded at nearly the $3\sigma$ level.
The results estimate the significance with the $s/\sqrt{b}$ estimator,
taking only statistical uncertainties under consideration.

\subsection{Side band analysis}
\label{subsec:sideband}

Estimating the background rates and distributions reliably is a very
challenging task, as uncertainties originating from the use of Monte
Carlo event generators and other theoretical calculations are often
too large. An alternative method that can work reasonably well requires an alteration of the selection setup, maximising
the background. The background shape and rate can then be modelled in
a region where the signal has little or no effect and subsequently extrapolated to the signal region. With this purpose in mind, the Shower
Deconstruction selection for the leading fat jet is loosened to obtain
a much lower efficiency of 26\% and no mass window is applied using the
BDRS tagger for the sub-leading fat jet.

\begin{figure}[!htb]
  \begin{center}
    \vspace*{3ex}
    \includegraphics[width=0.7\linewidth]{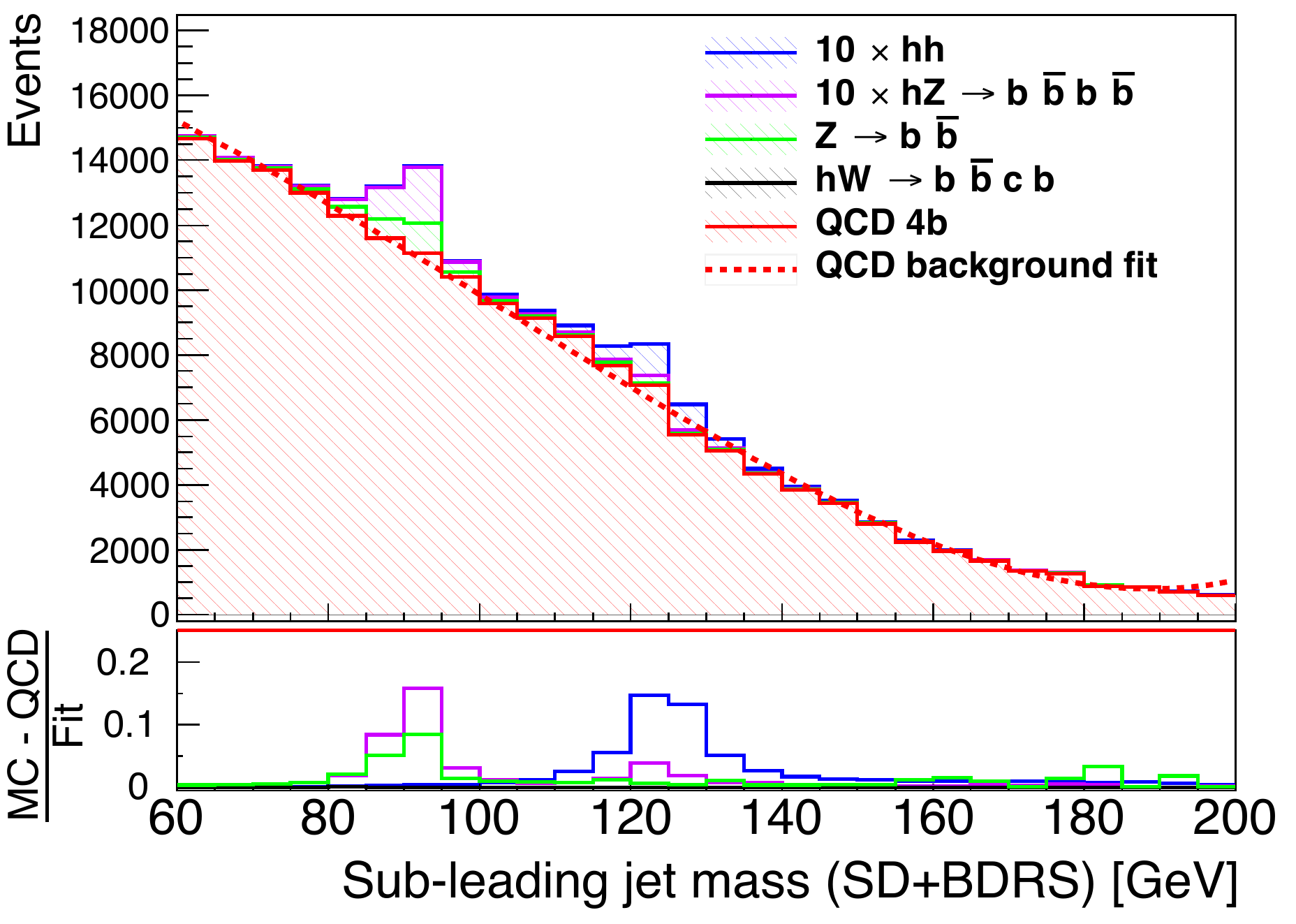}
    \vspace{-2ex}
  \end{center}
  \caption{A fit of a side band region using a 5$^\mathrm{th}$-order polynomial, performed with looser selection requirements, using Shower Deconstruction for the leading-$p_T$ Higgs boson identification and BDRS for the sub-leading Higgs mass reconstruction.
  The bottom part of the plot shows the different
  samples independently normalised by the fit function, to show the relative contribution of each one.}
  \label{fig:sideband}
\end{figure}

The
BDRS-reconstructed mass of the sub-leading jet is shown in
Fig.~\ref{fig:sideband}, including a model for the
leading QCD background using a 5$^{\textrm{th}}$-order polynomial, shown in
the dashed line. The ratio of the remaining backgrounds to the QCD $4b$
fit model is shown in the lower part of the plot with the $hh$ signal
and the $hZ$ background weighted by a factor of ten so that they can be compared.

The background estimate can be done by excluding the signal mass window
and the $Z$ boson mass region for the QCD $4b$ background model and using
the fit to estimate this background's content in the signal region. Other
significant backgrounds can be estimated in Monte Carlo simulation and
subtracted. These results can be transferred to the signal region
with a tighter Shower Deconstruction configuration.

It is interesting to point out that a comparison of the $hh
\rightarrow (b\bar{b}) (b\bar{b})$ signal, which varies with the
self-coupling $\lambda$, with the $hZ \rightarrow
(b\bar{b}) (b\bar{b})$ background as they both appear in
Fig.~\ref{fig:sideband}, can be used to estimate the self-coupling as a function of the Higgs-$Z$ coupling.

\section{Conclusions}
\label{sec:conclusions}
Using the BDRS method and Shower Deconstruction on the $(b\bar{b})(b\bar{b})$ final state in Higgs boson
  pair production, we have demonstrated that it is possible to obtain meaningful constraints on the Higgs
  boson triple self-coupling. 

  We considered explicitly the main
  irreducible backgrounds to our signal, including QCD $b\bar{b}b\bar{b}$,
  electroweak/QCD $Zb\bar{b}$ with $Z\rightarrow b\bar{b}$ and electroweak
  $hZ \rightarrow (b\bar{b})(b\bar{b})$, as well as the
  reducible $hW$ background containing charm-quark jets mis-tagged as
  $b$ jets. Including reducible backgrounds will not significantly
  alter our conclusions after requiring 4 $b$-tags. Moreover, we have demonstrated that a side-band analysis using Shower
  Deconstruction for the leading Higgs boson jet and BDRS for the
  sub-leading Higgs mass reconstruction can be a viable option in measuring the Higgs boson self-coupling.

We conclude that at an LHC running at 14~TeV, with 3000~fb$^{-1}$ of integrated luminosity,
  the self-coupling can be constrained to $\lambda \lesssim 1.2 \times
  \lambda_{SM}$ at 95\% confidence level based on statistical considerations alone, since no theoretical uncertainties have been included.
Nevertheless, using the side-band analysis proposed one could
estimate the background to good accuracy by extrapolating
the background content from real data in the side band into the
signal region. 

Further improvements are possible: refined trigger and $b$-tagging strategies can help to retain more signal and allow novel reconstruction techniques to achieve a better performance. These conclusions motivate in-depth examination of the $(b\bar{b})(b\bar{b})$ final state in Higgs boson by the LHC experimental collaborations. 

\appendix
\acknowledgments
AP would like to thank Paolo Torrielli for
useful discussion and acknowledges supported in part by the Swiss
National Science Foundation (SNF) under contract 200020-149517 and by
the European Commission through the ``LHCPhenoNet'' Initial Training
Network PITN-GA-2010-264564 and MCnetITN FP7 Marie Curie Initial Training Network
PITN-GA-2012-315877.
\bibliography{hh4b.bib}

\providecommand{\href}[2]{#2}\begingroup\raggedright\begin{thebibliography}{10}

\bibitem{ATLAS_Higgs}
{\bf ATLAS} Collaboration, G.~Aad et~al., {\it {Observation of a new particle
  in the search for the Standard Model Higgs boson with the ATLAS detector at
  the LHC}},  {\em Phys.Lett.} {\bf B716} (2012) 1--29,
  [\href{http://xxx.lanl.gov/abs/1207.7214}{{\tt arXiv:1207.7214}}].

\bibitem{CMS_Higgs}
{\bf CMS} Collaboration, S.~Chatrchyan et~al., {\it {Observation of a new boson
  at a mass of 125 GeV with the CMS experiment at the LHC}},  {\em Phys.Lett.}
  {\bf B716} (2012), no.~CMS-HIG-12-028, CERN-PH-EP-2012-220 30--61,
  [\href{http://xxx.lanl.gov/abs/1207.7235}{{\tt arXiv:1207.7235}}].

\bibitem{CMS-PAS-HIG-12-045}
{\bf CMS} Collaboration, {\it {Combination of standard model Higgs boson
  searches and measurements of the properties of the new boson with a mass near
  125 GeV}},  Tech. Rep. CMS-PAS-HIG-12-045, CERN, 2012.

\bibitem{ATLAS:2012wma}
{\bf ATLAS} Collaboration, {\it {Coupling properties of the new Higgs-like
  boson observed with the ATLAS detector at the LHC}},  Tech. Rep.
  ATLAS-CONF-2012-127, ATLAS-COM-CONF-2012-161, CERN, 2012.

\bibitem{ATLAS-CONF-2012-170}
{\bf ATLAS} Collaboration, {\it {An update of combined measurements of the new
  Higgs-like boson with high mass resolution channels}},  Tech. Rep.
  ATLAS-CONF-2012-170, CERN, Geneva, Dec, 2012.

\bibitem{Kanemura:2002vm}
S.~Kanemura, S.~Kiyoura, Y.~Okada, E.~Senaha, and C.~Yuan, {\it {New physics
  effect on the Higgs selfcoupling}},  {\em Phys.Lett.} {\bf B558} (2003)
  157--164, [\href{http://xxx.lanl.gov/abs/hep-ph/0211308}{{\tt
  hep-ph/0211308}}].

\bibitem{Grigo:2013rya}
J.~Grigo, J.~Hoff, K.~Melnikov, and M.~Steinhauser, {\it {On the Higgs boson
  pair production at the LHC}},  {\em Nucl.Phys.} {\bf B875} (2013) 1--17,
  [\href{http://xxx.lanl.gov/abs/1305.7340}{{\tt arXiv:1305.7340}}].

\bibitem{Plehn:2005nk}
T.~Plehn and M.~Rauch, {\it {The quartic higgs coupling at hadron colliders}},
  {\em Phys.Rev.} {\bf D72} (2005) 053008,
  [\href{http://xxx.lanl.gov/abs/hep-ph/0507321}{{\tt hep-ph/0507321}}].

\bibitem{Binoth:2006ym}
T.~Binoth, S.~Karg, N.~Kauer, and R.~Ruckl, {\it {Multi-Higgs boson production
  in the Standard Model and beyond}},  {\em Phys.Rev.} {\bf D74} (2006) 113008,
  [\href{http://xxx.lanl.gov/abs/hep-ph/0608057}{{\tt hep-ph/0608057}}].

\bibitem{Miller:1999ct}
D.~Miller and S.~Moretti, {\it {Can the trilinear Higgs selfcoupling be
  measured at future linear colliders?}},  {\em Eur.Phys.J.} {\bf C13} (2000)
  459--470, [\href{http://xxx.lanl.gov/abs/hep-ph/9906395}{{\tt
  hep-ph/9906395}}].

\bibitem{Glover:1987nx}
E.~N. Glover and J.~van~der Bij, {\it {Higgs boson pair production via gluon
  fusion}},  {\em Nucl.Phys.} {\bf B309} (1988) 282.

\bibitem{Dawson:1998py}
S.~Dawson, S.~Dittmaier, and M.~Spira, {\it {Neutral Higgs boson pair
  production at hadron colliders: QCD corrections}},  {\em Phys.Rev.} {\bf D58}
  (1998) 115012, [\href{http://xxx.lanl.gov/abs/hep-ph/9805244}{{\tt
  hep-ph/9805244}}].

\bibitem{Djouadi:1999rca}
A.~Djouadi, W.~Kilian, M.~Muhlleitner, and P.~Zerwas, {\it {Production of
  neutral Higgs boson pairs at LHC}},  {\em Eur.Phys.J.} {\bf C10} (1999)
  45--49, [\href{http://xxx.lanl.gov/abs/hep-ph/9904287}{{\tt
  hep-ph/9904287}}].

\bibitem{Plehn:1996wb}
T.~Plehn, M.~Spira, and P.~Zerwas, {\it {Pair production of neutral Higgs
  particles in gluon-gluon collisions}},  {\em Nucl.Phys.} {\bf B479} (1996)
  46--64, [\href{http://xxx.lanl.gov/abs/hep-ph/9603205}{{\tt
  hep-ph/9603205}}].

\bibitem{Baur:2002qd}
U.~Baur, T.~Plehn, and D.~L. Rainwater, {\it {Determining the Higgs boson
  selfcoupling at hadron colliders}},  {\em Phys.Rev.} {\bf D67} (2003) 033003,
  [\href{http://xxx.lanl.gov/abs/hep-ph/0211224}{{\tt hep-ph/0211224}}].

\bibitem{Baur:2003gp}
U.~Baur, T.~Plehn, and D.~L. Rainwater, {\it {Probing the Higgs selfcoupling at
  hadron colliders using rare decays}},  {\em Phys.Rev.} {\bf D69} (2004)
  053004, [\href{http://xxx.lanl.gov/abs/hep-ph/0310056}{{\tt
  hep-ph/0310056}}].

\bibitem{Dolan:2012rv}
M.~J. Dolan, C.~Englert, and M.~Spannowsky, {\it {Higgs self-coupling
  measurements at the LHC}},  {\em JHEP} {\bf 1210} (2012) 112,
  [\href{http://xxx.lanl.gov/abs/1206.5001}{{\tt arXiv:1206.5001}}].

\bibitem{Baglio:2012np}
J.~Baglio, A.~Djouadi, R.~Grober, M.~Muhlleitner, J.~Quevillon, et~al., {\it
  {The measurement of the Higgs self-coupling at the LHC: theoretical status}},
   \href{http://xxx.lanl.gov/abs/1212.5581}{{\tt arXiv:1212.5581}}.

\bibitem{Barr:2013tda}
A.~J. Barr, M.~J. Dolan, C.~Englert, and M.~Spannowsky, {\it {Di-Higgs final
  states augMT2ed -- selecting $hh$ events at the high luminosity LHC}},
  \href{http://xxx.lanl.gov/abs/1309.6318}{{\tt arXiv:1309.6318}}.

\bibitem{Dolan:2013rja}
M.~J. Dolan, C.~Englert, N.~Greiner, and M.~Spannowsky, {\it {Further on up the
  road: $hhjj$ production at the LHC}},
  \href{http://xxx.lanl.gov/abs/1310.1084}{{\tt arXiv:1310.1084}}.

\bibitem{Papaefstathiou:2012qe}
A.~Papaefstathiou, L.~L. Yang, and J.~Zurita, {\it {Higgs boson pair production
  at the LHC in the $b \bar{b} W^+ W^-$ channel}},
  \href{http://xxx.lanl.gov/abs/1209.1489}{{\tt arXiv:1209.1489}}.

\bibitem{Goertz:2013kp}
F.~Goertz, A.~Papaefstathiou, L.~L. Yang, and J.~Zurita, {\it {Higgs Boson
  self-coupling measurements using ratios of cross sections}},
  \href{http://xxx.lanl.gov/abs/1301.3492}{{\tt arXiv:1301.3492}}.

\bibitem{Goertz:2013eka}
F.~Goertz, A.~Papaefstathiou, L.~L. Yang, and J.~Zurita, {\it {Measuring the
  Higgs boson self-coupling at the LHC using ratios of cross sections}},
  \href{http://xxx.lanl.gov/abs/1309.3805}{{\tt arXiv:1309.3805}}.

\bibitem{deFlorian:2013jea}
D.~de~Florian and J.~Mazzitelli, {\it {Higgs pair production at NNLO}},
  \href{http://xxx.lanl.gov/abs/1309.6594}{{\tt arXiv:1309.6594}}.

\bibitem{deFlorian:2013uza}
D.~de~Florian and J.~Mazzitelli, {\it {Two-loop virtual corrections to Higgs
  pair production}},  {\em Phys.Lett.} {\bf B724} (2013) 306--309,
  [\href{http://xxx.lanl.gov/abs/1305.5206}{{\tt arXiv:1305.5206}}].

\bibitem{Cao:2013si}
J.~Cao, Z.~Heng, L.~Shang, P.~Wan, and J.~M. Yang, {\it {Pair Production of a
  125 GeV Higgs Boson in MSSM and NMSSM at the LHC}},  {\em JHEP} {\bf 1304}
  (2013) 134, [\href{http://xxx.lanl.gov/abs/1301.6437}{{\tt
  arXiv:1301.6437}}].

\bibitem{Gupta:2013zza}
R.~S. Gupta, H.~Rzehak, and J.~D. Wells, {\it {How well do we need to measure
  the Higgs boson mass and self-coupling?}},  {\em Phys.Rev.} {\bf D88} (2013)
  055024, [\href{http://xxx.lanl.gov/abs/1305.6397}{{\tt arXiv:1305.6397}}].

\bibitem{Nhung:2013lpa}
D.~T. Nhung, M.~Muhlleitner, J.~Streicher, and K.~Walz, {\it {Higher Order
  Corrections to the Trilinear Higgs Self-Couplings in the Real NMSSM}},
  \href{http://xxx.lanl.gov/abs/1306.3926}{{\tt arXiv:1306.3926}}.

\bibitem{Ellwanger:2013ova}
U.~Ellwanger, {\it {Higgs pair production in the NMSSM at the LHC}},  {\em
  JHEP} {\bf 1308} (2013) 077, [\href{http://xxx.lanl.gov/abs/1306.5541}{{\tt
  arXiv:1306.5541}}].

\bibitem{No:2013wsa}
J.~M. No and M.~Ramsey-Musolf, {\it {Probing the Higgs Portal at the LHC
  Through Resonant di-Higgs Production}},
  \href{http://xxx.lanl.gov/abs/1310.6035}{{\tt arXiv:1310.6035}}.

\bibitem{McCullough:2013rea}
M.~McCullough, {\it {A New Indirect Probe of the Higgs Self-Coupling}},
  \href{http://xxx.lanl.gov/abs/1312.3322}{{\tt arXiv:1312.3322}}.

\bibitem{Maierhofer:2013sha}
P.~Maierh{\"o}efer and A.~Papaefstathiou, {\it {Higgs Boson pair production
  merged to one jet}},  \href{http://xxx.lanl.gov/abs/1401.0007}{{\tt
  arXiv:1401.0007}}.

\bibitem{Hollik:2001px}
W.~Hollik and S.~Penaranda, {\it {Yukawa coupling quantum corrections to the
  selfcouplings of the lightest MSSM Higgs boson}},  {\em Eur.Phys.J.} {\bf
  C23} (2002) 163--172, [\href{http://xxx.lanl.gov/abs/hep-ph/0108245}{{\tt
  hep-ph/0108245}}].

\bibitem{Dubinin:1998nt}
M.~Dubinin and A.~Semenov, {\it {Triple and quartic interactions of Higgs
  bosons in the general two Higgs doublet model}},
  \href{http://xxx.lanl.gov/abs/hep-ph/9812246}{{\tt hep-ph/9812246}}.

\bibitem{Tian:2013yda}
J.~Tian and K.~Fujii, {\it {Measurement of Higgs couplings and self-coupling at
  the ILC}},  \href{http://xxx.lanl.gov/abs/1311.6528}{{\tt arXiv:1311.6528}}.

\bibitem{Dawson:2013bba}
S.~Dawson, A.~Gritsan, H.~Logan, J.~Qian, C.~Tully, et~al., {\it {Working Group
  Report: Higgs Boson}},  \href{http://xxx.lanl.gov/abs/1310.8361}{{\tt
  arXiv:1310.8361}}.

\bibitem{Lafaye:2000ec}
R.~Lafaye, D.~Miller, M.~Muhlleitner, and S.~Moretti, {\it {Double Higgs
  production at TeV colliders in the minimal supersymmetric standard model}},
  \href{http://xxx.lanl.gov/abs/hep-ph/0002238}{{\tt hep-ph/0002238}}.

\bibitem{Osland:1999ae}
P.~Osland and P.~Pandita, {\it {Multiple Higgs production and measurement of
  Higgs trilinear couplings in the MSSM}},
  \href{http://xxx.lanl.gov/abs/hep-ph/9911295}{{\tt hep-ph/9911295}}.

\bibitem{Osland:1999ad}
P.~Osland, {\it {Higgs boson production in e+ e- and e- e- collisions}},  {\em
  Acta Phys.Polon.} {\bf B30} (1999) 1967--1984,
  [\href{http://xxx.lanl.gov/abs/hep-ph/9903301}{{\tt hep-ph/9903301}}].

\bibitem{Brucherseifer:2013qva}
M.~Brucherseifer, R.~Gavin, and M.~Spira, {\it {MSSM Higgs Self-Couplings:
  Two-Loop $\mathcal{O}(\alpha_t \alpha_s)$ Corrections}},
  \href{http://xxx.lanl.gov/abs/1309.3140}{{\tt arXiv:1309.3140}}.

\bibitem{Yao:2013ika}
W.~Yao, {\it {Studies of measuring Higgs self-coupling with $HH\rightarrow
  b\bar b \gamma\gamma$ at the future hadron colliders}},
  \href{http://xxx.lanl.gov/abs/1308.6302}{{\tt arXiv:1308.6302}}.

\bibitem{Frederix:2014hta}
R.~Frederix, S.~Frixione, V.~Hirschi, F.~Maltoni, O.~Mattelaer, et~al., {\it
  {Higgs pair production at the LHC with NLO and parton-shower effects}},  {\em
  Phys.Lett.} {\bf B732} (2014) 142--149,
  [\href{http://xxx.lanl.gov/abs/1401.7340}{{\tt arXiv:1401.7340}}].

\bibitem{Seymour:1993mx}
M.~H. Seymour, {\it {Searches for new particles using cone and cluster jet
  algorithms: A Comparative study}},  {\em Z.Phys.} {\bf C62} (1994) 127--138.

\bibitem{Baur:2003gpa}
U.~Baur, T.~Plehn, and D.~L. Rainwater, {\it {Examining the Higgs boson
  potential at lepton and hadron colliders: A Comparative analysis}},  {\em
  Phys.Rev.} {\bf D68} (2003) 033001,
  [\href{http://xxx.lanl.gov/abs/hep-ph/0304015}{{\tt hep-ph/0304015}}].

\bibitem{Dolan:2012ac}
M.~J. Dolan, C.~Englert, and M.~Spannowsky, {\it {New Physics in LHC Higgs
  boson pair production}},  \href{http://xxx.lanl.gov/abs/1210.8166}{{\tt
  arXiv:1210.8166}}.

\bibitem{Gouzevitch:2013qca}
M.~Gouzevitch, A.~Oliveira, J.~Rojo, R.~Rosenfeld, G.~P. Salam, et~al., {\it
  {Scale-invariant resonance tagging in multijet events and new physics in
  Higgs pair production}},  {\em JHEP} {\bf 1307} (2013) 148,
  [\href{http://xxx.lanl.gov/abs/1303.6636}{{\tt arXiv:1303.6636}}].

\bibitem{Efrati:2014uta}
A.~Efrati and Y.~Nir, {\it {What if $\lambda_{hhh}\neq 3m_h^2/v$}},
  \href{http://xxx.lanl.gov/abs/1401.0935}{{\tt arXiv:1401.0935}}.

\bibitem{ATLAS-CONF-2014-005}
{\it {A search for resonant Higgs-pair production in the $b\bar{b}b\bar{b}$
  final state in $pp$ collisions at $\sqrt{s}=8$ TeV}},  Tech. Rep.
  ATLAS-CONF-2014-005, CERN, Geneva, Mar, 2014.

\bibitem{Cooper:2013kia}
B.~Cooper, N.~Konstantinidis, L.~Lambourne, and D.~Wardrope, {\it {Boosted $hh
  \rightarrow b\bar{b}b\bar{b}$: a new topology in searches for TeV-scale
  resonances at the LHC}},  {\em Phys.Rev.} {\bf D88} (2013) 114005,
  [\href{http://xxx.lanl.gov/abs/1307.0407}{{\tt arXiv:1307.0407}}].

\bibitem{Han:2013sga}
C.~Han, X.~Ji, L.~Wu, P.~Wu, and J.~M. Yang, {\it {Higgs pair production with
  SUSY QCD correction: revisited under current experimental constraints}},
  {\em JHEP} {\bf 1404} (2014) 003,
  [\href{http://xxx.lanl.gov/abs/1307.3790}{{\tt arXiv:1307.3790}}].

\bibitem{Shao:2013bz}
D.~Y. Shao, C.~S. Li, H.~T. Li, and J.~Wang, {\it {Threshold resummation
  effects in Higgs boson pair production at the LHC}},
  \href{http://xxx.lanl.gov/abs/1301.1245}{{\tt arXiv:1301.1245}}.

\bibitem{Philippov:2006th}
Y.~Philippov, {\it {Yukawa radiative corrections to the trilinear
  self-couplings of neutral CP-even Higgs bosons and decay width Gamma (H
  $\rightarrow$ hh) in the MSSM}},  {\em Phys.Atom.Nucl.} {\bf 70} (2007)
  1288--1293, [\href{http://xxx.lanl.gov/abs/hep-ph/0611260}{{\tt
  hep-ph/0611260}}].

\bibitem{Dolgopolov:2003kv}
M.~Dolgopolov and Y.~Philippov, {\it {The Trilinear neutral Higgs selfcouplings
  in the MSSM: Complete one loop analysis}},
  \href{http://xxx.lanl.gov/abs/hep-ph/0310018}{{\tt hep-ph/0310018}}.

\bibitem{Boudjema:2001ii}
F.~Boudjema and A.~Semenov, {\it {Measurements of the SUSY Higgs selfcouplings
  and the reconstruction of the Higgs potential}},  {\em Phys.Rev.} {\bf D66}
  (2002) 095007, [\href{http://xxx.lanl.gov/abs/hep-ph/0201219}{{\tt
  hep-ph/0201219}}].

\bibitem{Butterworth:2008iy}
J.~M. Butterworth, A.~R. Davison, M.~Rubin, and G.~P. Salam, {\it {Jet
  substructure as a new Higgs search channel at the LHC}},  {\em
  Phys.Rev.Lett.} {\bf 100} (2008) 242001,
  [\href{http://xxx.lanl.gov/abs/0802.2470}{{\tt arXiv:0802.2470}}].

\bibitem{Soper:2011cr}
D.~E. Soper and M.~Spannowsky, {\it {Finding physics signals with shower
  deconstruction}},  {\em Phys.Rev.} {\bf D84} (2011) 074002,
  [\href{http://xxx.lanl.gov/abs/1102.3480}{{\tt arXiv:1102.3480}}].

\bibitem{Soper:2012pb}
D.~E. Soper and M.~Spannowsky, {\it {Finding top quarks with shower
  deconstruction}},  {\em Phys.Rev.} {\bf D87} (2013), no.~5 054012,
  [\href{http://xxx.lanl.gov/abs/1211.3140}{{\tt arXiv:1211.3140}}].

\bibitem{Soper:2014rya}
D.~E. Soper and M.~Spannowsky, {\it {Finding physics signals with event
  deconstruction}},  \href{http://xxx.lanl.gov/abs/1402.1189}{{\tt
  arXiv:1402.1189}}.

\bibitem{Bartoldus:1602235}
R.~S. Bartoldus, C.~M.~C. Bee, D.~C. Francis, N.~R. Gee, S.~L.~R. George,
  R.~M.~S. Hauser, R.~R. Middleton, T.~C. Pauly, O.~K. Sasaki, D.~O. Strom,
  R.~R.~I. Vari, and S.~R.~I. Veneziano, {\it {Technical Design Report for the
  Phase-I Upgrade of the ATLAS TDAQ System}},  Tech. Rep. CERN-LHCC-2013-018.
  ATLAS-TDR-023, CERN, Geneva, Sep, 2013.
\newblock Final version presented to December 2013 LHCC.

\bibitem{ATLAS_btagging}
T.~A. collaboration, {\it {Search for the Standard Model Higgs boson produced
  in association with top quarks and decaying to $b\bar{b}$ in pp collisions at
  $\sqrt{s} =$ 8 TeV with the ATLAS detector at the LHC}}, .

\bibitem{Thaler:2010tr}
J.~Thaler and K.~Van~Tilburg, {\it {Identifying Boosted Objects with
  N-subjettiness}},  {\em JHEP} {\bf 1103} (2011) 015,
  [\href{http://xxx.lanl.gov/abs/1011.2268}{{\tt arXiv:1011.2268}}].

\bibitem{Soper:2010xk}
D.~E. Soper and M.~Spannowsky, {\it {Combining subjet algorithms to enhance ZH
  detection at the LHC}},  {\em JHEP} {\bf 1008} (2010) 029,
  [\href{http://xxx.lanl.gov/abs/1005.0417}{{\tt arXiv:1005.0417}}].

\bibitem{Backovic:2012jj}
M.~Backovic, J.~Juknevich, and G.~Perez, {\it {Boosting the Standard Model
  Higgs Signal with the Template Overlap Method}},  {\em JHEP} {\bf 1307}
  (2013) 114, [\href{http://xxx.lanl.gov/abs/1212.2977}{{\tt
  arXiv:1212.2977}}].

\bibitem{Almeida:2011aa}
L.~G. Almeida, O.~Erdogan, J.~Juknevich, S.~J. Lee, G.~Perez, et~al., {\it
  {Three-particle templates for a boosted Higgs boson}},  {\em Phys.Rev.} {\bf
  D85} (2012) 114046, [\href{http://xxx.lanl.gov/abs/1112.1957}{{\tt
  arXiv:1112.1957}}].

\bibitem{Ellis:2012sn}
S.~D. Ellis, A.~Hornig, T.~S. Roy, D.~Krohn, and M.~D. Schwartz, {\it {Qjets: A
  Non-Deterministic Approach to Tree-Based Jet Substructure}},  {\em
  Phys.Rev.Lett.} {\bf 108} (2012) 182003,
  [\href{http://xxx.lanl.gov/abs/1201.1914}{{\tt arXiv:1201.1914}}].

\bibitem{Bahr:2008pv}
M.~Bahr, S.~Gieseke, M.~Gigg, D.~Grellscheid, K.~Hamilton, et~al., {\it
  {Herwig++ Physics and Manual}},  {\em Eur.Phys.J.} {\bf C58} (2008) 639--707,
  [\href{http://xxx.lanl.gov/abs/0803.0883}{{\tt arXiv:0803.0883}}].

\bibitem{Arnold:2012fq}
K.~Arnold, L.~d'Errico, S.~Gieseke, D.~Grellscheid, K.~Hamilton, et~al., {\it
  {Herwig++ 2.6 Release Note}},  \href{http://xxx.lanl.gov/abs/1205.4902}{{\tt
  arXiv:1205.4902}}.

\bibitem{Bellm:2013lba}
J.~Bellm, S.~Gieseke, D.~Grellscheid, A.~Papaefstathiou, S.~Platzer, et~al.,
  {\it {Herwig++ 2.7 Release Note}},
  \href{http://xxx.lanl.gov/abs/1310.6877}{{\tt arXiv:1310.6877}}.

\bibitem{hpair}
``hpair program.'' {http://people.web.psi.ch/spira/hpair/}.

\bibitem{Mangano:2002ea}
M.~L. Mangano, M.~Moretti, F.~Piccinini, R.~Pittau, and A.~D. Polosa, {\it
  {ALPGEN, a generator for hard multiparton processes in hadronic collisions}},
   {\em JHEP} {\bf 0307} (2003) 001,
  [\href{http://xxx.lanl.gov/abs/hep-ph/0206293}{{\tt hep-ph/0206293}}].

\bibitem{Frixione:2010ra}
S.~Frixione, F.~Stoeckli, P.~Torrielli, and B.~R. Webber, {\it {NLO QCD
  corrections in Herwig++ with MC@NLO}},  {\em JHEP} {\bf 1101} (2011) 053,
  [\href{http://xxx.lanl.gov/abs/1010.0568}{{\tt arXiv:1010.0568}}].

\bibitem{Alwall:2011uj}
J.~Alwall, M.~Herquet, F.~Maltoni, O.~Mattelaer, and T.~Stelzer, {\it {MadGraph
  5 : Going Beyond}},  {\em JHEP} {\bf 1106} (2011) 128,
  [\href{http://xxx.lanl.gov/abs/1106.0522}{{\tt arXiv:1106.0522}}].

\bibitem{Alwall:2014hca}
J.~Alwall, R.~Frederix, S.~Frixione, V.~Hirschi, F.~Maltoni, et~al., {\it {The
  automated computation of tree-level and next-to-leading order differential
  cross sections, and their matching to parton shower simulations}},
  \href{http://xxx.lanl.gov/abs/1405.0301}{{\tt arXiv:1405.0301}}.

\bibitem{Binoth:2009rv}
T.~Binoth, N.~Greiner, A.~Guffanti, J.~Reuter, J.-P. Guillet, et~al., {\it
  {Next-to-leading order QCD corrections to pp $\rightarrow$ b anti-b b anti-b
  + X at the LHC: the quark induced case}},  {\em Phys.Lett.} {\bf B685} (2010)
  293--296, [\href{http://xxx.lanl.gov/abs/0910.4379}{{\tt arXiv:0910.4379}}].

\bibitem{Greiner:2011mp}
N.~Greiner, A.~Guffanti, T.~Reiter, and J.~Reuter, {\it {NLO QCD corrections to
  the production of two bottom-antibottom pairs at the LHC}},  {\em
  Phys.Rev.Lett.} {\bf 107} (2011) 102002,
  [\href{http://xxx.lanl.gov/abs/1105.3624}{{\tt arXiv:1105.3624}}].

\end{thebibliography}\endgroup
\bibliographystyle{JHEP.bst}

\end{document}